\documentclass[usenatbib,usegraphicx]{mn2e}
\pdfoutput=0
\usepackage{ifpdf}

\usepackage{amsmath}
\usepackage{graphicx}
\usepackage{color}
\usepackage{fixltx2e}
\DeclareGraphicsExtensions{.eps,.pdf,.png,.jpg}

\usepackage[]{hyperref}

\hypersetup{
    colorlinks=true,
    linkcolor=black,
    citecolor=black,
    filecolor=black,
    urlcolor=black,
}

\begin{document}

    \ifpdf
    \else

\title[Equilibrium in disk galaxies]{Balance among gravitational instability, star formation, and accretion determines the structure and evolution of disk galaxies}

\author[Forbes et al.]{John C. Forbes,$^1$\textsuperscript{\thanks{E-mail: jforbes@ucolick.org}} Mark R. Krumholz,$^1$ Andreas Burkert,$^{2,3}$\textsuperscript{\thanks{Max Planck Fellow}} Avishai Dekel$^4$ \\
$^1$Department of Astronomy \& Astrophysics, University of California, Santa Cruz, CA 95064 USA \\
$^2$University Observatory Munich (USM), Scheinerstrasse 1, 81679 Munich, Germany \\
$^3$Max-Planck-Institut fuer extraterrestrische Physik, Giessenbachstrasse 1, 85758 Garching, Germany \\
$^4$Racah Institute of Physics, The Hebrew University, Jerusalem 91904 Israel
}

\maketitle

\begin{abstract}
Over the past 10 Gyr, star-forming galaxies have changed dramatically, from clumpy and gas rich, to rather quiescent stellar-dominated disks with specific star formation rates lower by factors of a few tens. We present a general theoretical model for how this transition occurs, and what physical processes drive it, making use of 1D axisymmetric thin disk simulations with an improved version of the Gravitational Instability-Dominated Galaxy Evolution Tool (GIDGET) code. We show that at every radius galaxies tend to be in a slowly evolving equilibrium state wherein new accretion is balanced by star formation, galactic winds, and radial transport of gas through the disk by gravitational instability (GI) -driven torques. The gas surface density profile is determined by which of these terms are in balance at a given radius, - direct accretion is balanced by star formation and galactic winds near galactic centers, and by transport at larger radii. We predict that galaxies undergo a smooth transition from a violent disk instability phase to secular evolution. This model provides a natural explanation for the high velocity dispersions and large clumps in $z\sim 2$ galaxies, the growth and subsequent quenching of bulges, and features of the neutral gas profiles of local spiral galaxies.
\end{abstract}


\section{Introduction}

Historically astronomers have studied the evolution of galaxies through changes in their stellar populations. The real action, though, takes place in the gas phase. However, it is only recently that observations in the radio have had sufficient sensitivity to detect molecular gas in emission at high redshift, and sufficient resolution to map both molecular and atomic gas in great detail for nearby galaxies. Integral field and grism spectroscopy of H$\alpha$ have also opened a new view on the spatial distribution of star formation and gas kinematics at $z \sim 1-2$.  

Numerous surveys have shown that the specific star formation rates (sSFR, the star formation rate divided by the stellar mass) of Milky Way (MW) mass galaxies have decreased by roughly a factor of 20 since $z=2$. With the wide acceptance of $\Lambda$CDM cosmology, which entails the hierarchical growth of dark matter haloes, it became common lore that mergers were a major driver of this dramatic change in the nature of galaxies. More recently though, the small scatter in the correlation between the stellar mass $M_*$ and the star formation rate (the star forming main sequence) for galaxies out to $z=2$ has suggested that most stellar mass growth occurs in galaxies that are not undergoing dramatic merger events, but rather in typical-looking disks \citep[e.g.][]{Noeske2007,Rodighiero2011,Kaviraj2013}. Maps of H$\alpha$ emission in main sequence galaxies confirm that star formation occurs in radially extended disks at $z\sim 1$ \citep{Nelson2013}.

Even though the higher star formation rates at $z\sim 2$ are unlikely to be caused by mergers, galaxies where the sSFR's are so much higher than in local galaxies must be dramatically different. This has been verified directly by gas-phase observations, which show that these galaxies are gas-rich \citep{Tacconi2010a,Tacconi2012}, highly turbulent \citep{Cresci2009,ForsterSchreiber2009}, and gravitationally unstable \citep{Burkert2010,Genzel2011}. These differences are also reflected in the optical morphologies, which are distinctly clumpy \citep{Elmegreen2004,Elmegreen2005}.

High resolution hydrodynamical simulations \citep{Bournaud2009,Ceverino2010} have strongly suggested that the reason these galaxies are so different from low-redshift disks is rapid gas accretion from the cosmic web through cold dense filaments \citep{Dekel2009}, which in turn leads to galaxies with low values of the Toomre $Q$ parameter \citep{Toomre1964},
\begin{equation}
Q_\mathrm{Toomre} = \frac{\kappa \sigma_d}{\pi G \Sigma_d}
\end{equation}
Here $\kappa(r)=\sqrt{2(\beta(r)+1)}\Omega(r)$ is the epicyclic frequency, which is roughly comparable to the angular frequency $\Omega$, depending on the local powerlaw slope of the rotation curve, $\beta = d\ln v_\phi/d\ln r$. The velocity dispersion and surface density of the disk material are $\sigma_d$ and $\Sigma_d$ respectively. This instability has dramatic effects on the dynamics of the disk \citep*{Dekel2009a}. In regions where $Q_\mathrm{Toomre} \la 1$, the disk is unstable to axisymmetric perturbations on a scale $\lambda \sim \sigma_d^2/G\Sigma_d$, leading to clumps of this characteristic size. The clumpiness of the disk will in turn drive turbulence through the random torques exerted by the inhomogeneous gravitational field on material in the disk. The ultimate source of this kinetic energy is the gravitational potential of the galaxy, so mass must flow inwards \citep[e.g.][]{Gammie2001,Dekel2009a} (though some will flow outwards to conserve angular momentum). As a result of this the turbulent velocity dispersion $\sigma_d$, and hence $Q_\mathrm{Toomre}$, is increased, so given a sufficient gas supply, the value of $Q_\mathrm{Toomre}$ will be self-regulated to a marginally stable value of order unity.

Alternative scenarios for driving the turbulence and producing clumps have been explored by other authors. \citet{Genel2012a} constructed a simple model for the scenario in which the turbulence is driven by the kinetic energy of material as it accretes onto the disk \citep[see also][]{Elmegreen2010}. The details of the origin of the clumpy morphologies has also come under recent theoretical and observational investigation, and the importance of ex-situ clumps from minor mergers is not negligible (Mendelkar et al in prep). Supernovae \citep{Joung2009}, radiation pressure \citep{Krumholz2012a,Krumholz2013}, and the two working in tandem \citep{Agertz2012}, being the primary sources of energy outside of gravitational potential energy, have also been studied as drivers of turbulence and outflows.

Undoubtedly all of these processes occur. All of the sources of stellar feedback suffer from a great deal of uncertainty in the degree to which they couple with the interstellar medium, and typically require extremely high resolution hydrodynamical simulations to model properly. The highest resolution simulations to date, those of \citet{Krumholz2012a,Krumholz2013} for radiation pressure and those of \citet{Joung2009} for SNe, suggest that these sources of turbulence are unable to produce the high velocity dispersions observed in $z\sim 2$ disks. The gravitational instability scenario has the advantage that it is difficult to avoid; if $Q_\mathrm{Toomre}\la1$, gas will collapse and drive turbulence. Simple analytic arguments also suggest that the GI scenario leads to the correct behavior of $\sigma/v_{circ}$ over time, whereas the direct kinetic energy injection scenario does not \citep{Genel2012a}. Moreover, even $z=0$ disk galaxies have values of $Q_\mathrm{Toomre}$ (when corrected for multiple components and finite disk thickness) of order unity \citep{Romeo2011}.

In this work we build on the physical picture presented in simple toy models \citep{Dekel2009a,Cacciato2012} of the gravitational instability and how it evolves over time. \citet{Krumholz2010a} developed a formalism to show how gravitationally unstable disks behave as a function of radius in steady state and how quickly the disks approach steady state. In \citet*[][hereafter F12]{Forbes2012}, we extended the time-dependent numerical model of \citet{Krumholz2010a} to include star formation, stellar migration, and metallicity evolution to give a realistic picture for how galaxies evolve over cosmological times with all these processes. In this work, rather than focus on the stellar populations, we explore what sets the gas distribution. Our model includes a number of improvements over the models presented in F12 which we discuss in detail in appendix \ref{app:changes}, and a new stellar migration formalism (appendix \ref{app:stmig}).

One of our goals here is to understand the connection between the high redshift star forming galaxies and their $z=0$ descendants. The two galaxy populations are vastly different in terms of their gas fractions and sSFR's, yet remarkably similar in morphology. Recent $z=0$ measurements of the structure of gas in nearby spirals, The HI Nearby Galaxy Survey (THINGS) \citep{Walter2008} and the HERA CO-Line Extragalactic Survey (HERACLES) \citep{Leroy2009}, have provided unprecedented high spatial resolution data. These data have been fundamental in our understanding of star formation, and \citet{Bigiel2012} recently showed that these galaxies exhibit a universal gas surface density profile with remarkably small scatter. 

The general problem of how to connect high-redshift galaxy populations to their low-redshift counterparts has been approached for the past few decades with semi-analytical models (SAMs). These models are generally built on top of dark matter merger trees constructed from N-body cosmological simulations. Each galaxy is typically treated as a simple system described by a few quantities, e.g. cold and hot gas mass, stellar mass, black hole mass, and the entire population evolves according to parameterized recipes for gas cooling, star formation, stellar feedback, black hole growth, mergers, etc. With a few exceptions (\citet{vandenBosch2001} with subsequent work by \citet{Dutton2007, Dutton2009} and the simpler \citet{Fu2010}), SAMs have not tracked quantities as a function of radius (or more accurately specific angular momentum). The only model where matter can change its specific angular momentum \citep{Fu2013} does so in an ad-hoc way with no physical justification. This work attempts to fill this void without resorting to extremely expensive 3D hydrodynamical simulations, which must necessarily be either very low-resolution to see a large number of galaxies \citep[e.g.][]{Dave2011} or one galaxy at a time \citep[e.g.][]{Guedes2011}.

In section \ref{sec:code} we review the equations solved by our 1D code. The results of a wide range of simulations done with the code are presented in section \ref{sec:results}. We discuss the implications in section \ref{sec:discussion} and summarize in section \ref{sec:conclusion}.
	
\section{The GIDGET code}
\label{sec:code}

Our one-dimensional disk galaxy evolution code, GIDGET\footnote{The source code repository is freely available from \url{http://www.ucolick.org/~jforbes/gidget.html}}, is described in more detail in F12. The code tracks the surface density, velocity dispersion, and metallicity of one gas component and one or more stellar components, as a function of radius and time. The following subsections will describe the evolution equations for these quantities in some detail; we include a comprehensive list of all parameters used in this study, defined below, in table \ref{ta:params}. The most important physical ingredients are star formation, external accretion onto the disk, and radial transport of gas through the disk.

\subsection{Gas transport and cooling}
\label{sec:gasTransport}
GIDGET solves the full equations of hydrodynamics in the limit of a thin, axisymmetric, rotationally-supported disk, supported vertically by supersonic turbulent pressure. In this limit, the state of the gas at a particular time is described by a surface density $\Sigma(r)$ and a velocity dispersion $\sigma(r)=\sqrt{\sigma_{turb}^2(r)+\sigma_{th}^2}$ with a turbulent and thermal component.

\begin{table*}
        \begin{minipage}{150mm}
        \caption{An exhaustive list of all parameters used in this study}
        \label{ta:params}
        \begin{tabular}{cccl}
        \hline
        Parameter & Fiducial Value & Plausible Range & Description \\  
        \hline
        & & & {\bf Gas Migration} (section \ref{sec:gasTransport}) \\
        $\eta$ & 1.5 & 0.5 -- 4.5 & $(3/2)$ kinetic energy dissipation rate per scale-height crossing time \\
        $Q_{GI}$ & 2 & 1--3 & Marginally stable value of $Q$ \\
        $T_\mathrm{gas}$ & 7000 K & 3000--$10^4$& Gas temperature; sets the minimum gas velocity dispersion \\ 
        $\alpha_{MRI}$ & 0.01 & 0--0.1 & Value of $T_{r\phi}/\rho\sigma_{th}^2$ without gravitational instability \\ \hline
        & & & {\bf Rotation Curve} (section \ref{sec:rotCurve}) \\
        $v_\mathrm{circ}$ & 220 km s$^{-1}$ & 180--250 & Circular velocity in flat part of rotation curve\\
        $r_b$ & 3 kpc & 0--10 kpc & Radius where rotation curve transitions from powerlaw to flat \\
        $\beta_0$ & 0.5 & 0--1 & Powerlaw slope of $v_\phi(r)$ at small radii \\
        $n$ & 2 & 1--5 & Sharpness of the transition in the rotation curve \\ \hline
        & & & {\bf Star Formation} (section \ref{sec:SF}) \\
        $\epsilon_\mathrm{ff}$ & 0.01 & .003--.03 & Star formation efficiency per freefall time in the Toomre regime \\
        $f_{{\rm H}_2,min}$ & 0.03 & .01--.1 & Minimum $f_{{\rm H}_2}$. \\
        $t_{SC}$ & 2 Gyr & 1--3 Gyr & Depletion time of ${\rm H}_2$ in the single cloud regime \\
        $f_R$ & 0.54 & $.4+$ & Mass fraction of a zero-age stellar population not recycled to the ISM \\
        $\mu$ & 0.5 & 0--2 & Galactic winds' mass loading factor \\ \hline
        & & & {\bf Metallicity} (section \ref{sec:metallicity}) \\
        $y$ & .054 & .05--.07 & Mass of metals yielded per mass locked in stellar remnants \\
        $\xi$ & 0 & 0--1 & Metallicity enhancement of galactic winds \\
        $Z_{IGM}$ & $0.1 Z_\odot = 0.002$ & $(.01$--1$) Z_\odot$ & Metallicity of initial and infalling baryons \\
        $k_Z$ & 1 & .3--3 & Amplitude of metallicity diffusion relative to \citep{Yang2012}\\ \hline
        & & & {\bf Stellar Migration} (appendix \ref{app:stmig}) \\
        $Q_\mathrm{lim}$ & 2.5 & 2--3 & Value of $Q_*$ below which spiral instabilities will heat the stars \\
        $T_\mathrm{mig}$ & 4 & 2--5 & Number of local orbital times over which stars are heated by spiral instabilities \\ \hline
        & & & {\bf Accretion} (section \ref{sec:acc}) \\
        $M_{h,0}$ & $10^{12} M_\odot$ & - & Halo mass at $z=0$ \\
        $\Delta \omega$ & 0.5 & 0.1--1 & Interval of $\omega \sim z$ over which accretion rate is constant \\
        $r_\mathrm{acc}(z=0)$ & 6.9 kpc & 3--20 kpc & Scale length of new infalling gas \\
        $\beta_z$ & 0.38 & 0--1 & Scaling of efficiency with $(1+z)$\\
        $\beta_{M_h}$ & -0.25 & -1--0 & Scaling of efficiency with halo mass \\
        $\epsilon_0$ & 0.31 & $\sim 0$ - .5 & Efficiency at $M_h=10^{12} M_\odot$, $z=0$\\
        $\epsilon_{max}$ & 1 & 0.5--1 & Maximum value of efficiency\\ \hline
        & & & {\bf Initial Conditions} (section \ref{sec:IC}) \\
        $\alpha_r$ & 1/3 & 0--1 & Scaling of accretion scale length with halo mass \\
        $f_{g,0}$ & 0.5 & 0.2--0.7 &Initial gas fraction  \\
        $f_{cool}$ & 1 & 0.4--1 & Fraction of $f_b M_h(z=z_{relax})$ contained in the initial disk\\
        $z_{relax}$ & 2.5 & 2--3 & $z$ at which the simulation is initialized\\
        $\phi_0$ & 1 & 1--5 & Initial ratio of stellar to gaseous velocity dispersion\\ \hline
        & & & {\bf Computational Domain} (see F12) \\
        $x_0$ & .004 & $0<x_0\ll 1$ & Inner edge of domain as a fraction of $R$ \\
        $R$ & 40 kpc &10--100 kpc & Outer edge of domain \\
        $n_x$ & 200 & $\ga 100$ & Number of radial cells \\
        tol & $10^{-4}$ & $10^{-5}--x_0$ & Fastest change allowed in state variables, per orbital time at $r=R$\\ \hline
        & & & {\bf Cosmology}\footnote{We are restricted to using the WMAP5 cosmology because the stochastic accretion histories use fits to N-body simulations with those cosmological parameters.} (section \ref{sec:acc}) \\
        $\Omega_m$ & 0.258 & - & Average present-day matter energy density as a fraction of the critical density\\
        $1-\Omega_m-\Omega_\Lambda$ & 0 & - & Deviation from a flat universe \\
        $f_b$ & 0.17 & - & Universal baryon fraction \\
        $H_0$ & 72 km\ s$^{-1}$ Mpc$^{-1}$ & - & Hubble's constant \\
        $\sigma_8$ & .796 & - & Normalization of the dark matter power spectrum \\ \hline
        \end{tabular}
\end{minipage}
\end{table*}

The change in gas surface density at a given radius is described by a simple continuity equation accounting for mass flow through the disk, with source terms for star formation, recycling of gas by stellar mass loss, galactic winds, and cosmological accretion.
\begin{equation}
\label{eq:dcoldt}
\frac{\partial \Sigma}{\partial t} = \frac{1}{2\pi r}\frac{\partial}{\partial r} \dot{M} - (f_R + \mu)\dot{\Sigma}_*^{SF} + \dot{\Sigma}_{cos}.
\end{equation}
The first term represents the flow of mass within the disk, where $\dot{M}$ is defined as the net gas mass per unit time moving towards the center of the disk across cylindrical radius $r$. Typically $\dot{M}>0$, representing inward mass flux, but negative values at large radii in the disk are generally necessary to conserve angular momentum. The second term of the continuity equation represents gas forming stars. Only a fraction $f_R$ of that gas will remain in stellar remnants, while the remainder will be recycled to the ISM; we approximate this process as instantaneous as suggested by \citet{Tinsley1980}. Mass is also ejected at each radius in galactic scale winds in proportion to the star formation rate, with mass loading factor $\mu$. Finally, $\dot{\Sigma}_{cos}$ represents the rate of cosmological accretion onto the disk. The winds are assumed to escape the galaxy, though in principle they could be re-accreted later through this final term.

To evolve the velocity dispersion of the gas, we employ the energy equation added to the dot product of ${\bf v}$ with the momentum equation, yielding a total (kinetic + internal) energy equation,
 \begin{equation}
 \label{eq:dsigdt}
 \frac{\partial \sigma}{\partial t} = \frac{\mathcal{G}-\mathcal{L}}{3\sigma\Sigma} + \frac{\sigma}{6\pi r \Sigma}\frac{\partial}{\partial r} \dot{M} + \frac{5 (\partial\sigma/\partial r)}{6\pi r \Sigma}\dot{M} + \frac{(\beta-1)v_\phi}{6\pi r^3\Sigma \sigma}\mathcal{T}.
 \end{equation} 
Radiative gains and losses per unit area, respectively $\mathcal{G}$ and $\mathcal{L}$, are encompassed in the first term. The second and third terms account for the advection of kinetic energy as the gas moves through the disk. The torques which move gas radially in the disk, included in the final term, transfer energy between the galactic potential and the turbulent velocity dispersion. Here $\mathcal{T} = \int 2\pi r^2 T_{r\phi} dz$ is the vertically integrated effective viscous torque. Note that physically $T_{r\phi} \le 0$, and for rotation curves flatter than solid-body $\beta < 1$, so this final term adds kinetic energy to the gas.
 
The viscous torque is related to the mass flux via the conservation of angular momentum, as derived from the $\phi-$component of the Navier-Stokes equations:
\begin{equation}
\label{eq:angMom}
\dot{M} \equiv -2\pi r\Sigma v_r =  -\frac{1}{v_\phi (1+\beta)} \frac{\partial \mathcal{T}}{\partial r}.
\end{equation}
The mass flux, or equivalently the gas velocity in the radial direction, or equivalently the torque, are not known {\it ab initio}. To calculate them modelers have historically, since \citet{Shakura1973}, appealed to an order-of-magnitude argument, namely that $T_{r\phi} = \alpha \rho \sigma^2$, or equivalently $\nu = \alpha \sigma H$ where $\alpha$ is a parameter that might be measured from hydrodynamical simulations, $\nu$ is the resultant effective turbulent viscosity and $H$ is the scale height.  Physical causes for the turbulence include the magneto-rotational instability (MRI) and gravitational instability (GI). The value of $\alpha$ measured from simulations of the former varies by orders of magnitude, but is generally less than 0.1, particularly if the magnetic field is not forced to be vertical \citep{Balbus1998}. To distinguish between gravitational instability, which we model in a more consistent way, and the MRI or any other source of turbulence, which we include for comparison, we split our variables related to the torque into two components, $\mathcal{T} = \mathcal{T}_{GI} + \mathcal{T}_{MRI}$, and similarly for $v_r$, $T_{r\phi}$, $\dot{M}$, and $\alpha$ (the effects can just be added together since all of our equations are linear in these quantities).

Rather than pick a constant value of $\alpha_{GI}$, we calculate at every timestep the value of $\mathcal{T}_{GI}(r)$ such that in regions where $Q \le Q_{GI}$, the torques will act to move and heat the gas so that $dQ/dt = 0$. In regions of the disk where $Q > Q_{GI}$, $\mathcal{T}_{GI} = \dot{M}_{GI}=v_{r,GI} = 0$. To see how this works, consider the rate of change of $Q$ with time,
\begin{eqnarray}
\frac{dQ}{dt} &=& \frac{\partial \Sigma}{\partial t}\frac{\partial Q}{\partial \Sigma} + \frac{\partial \sigma}{\partial t}\frac{\partial Q}{\partial \sigma} + \nonumber \\
& &  \frac{\partial \Sigma_*}{\partial t}\frac{\partial Q}{\partial \Sigma_*} + \frac{\partial \sigma_{rr}}{\partial t}\frac{\partial Q}{\partial \sigma_{rr}} + \frac{\partial \sigma_{zz}}{\partial t}\frac{\partial Q}{\partial \sigma_{zz}}  \nonumber \\
 & = & f_\mathrm{transport}\left(\Sigma,\sigma,\Sigma_*,\sigma_{rr},\sigma_{zz},\mathcal{T}_{GI},\frac{\partial \mathcal{T}_{GI}}{\partial r}, \frac{\partial^2 \mathcal{T}_{GI}}{\partial r^2}\right) \nonumber  \\
 & &+ f_\mathrm{source}\left(\Sigma,\sigma,\Sigma_*,\sigma_{rr},\sigma_{zz}, Z  \right) .
\end{eqnarray}
The first equation is simply an application of the chain rule, while the second is just a definition, wherein we split all the terms into those which depend on $\mathcal{T}_{GI}$ and those which do not. Note that the source term includes the terms related to the $\alpha_{MRI}$-viscosity, star formation, and radiative cooling. The function $f_\mathrm{transport}$ has the nice property that it is linear in $\mathcal{T}_{GI}$ and its spatial derivatives, so when $\mathcal{T}_{GI}=0$, $f_\mathrm{transport}=0$ and we are left with $dQ/dt = f_\mathrm{source}$. Meanwhile in regions where $Q<Q_{GI}$ (by some small amount), we solve the equation $f_\mathrm{transport} = - f_\mathrm{source}$, i.e. we force $dQ/dt=0$. Because $f_\mathrm{transport}$ is linear, this equation may be solved efficiently for $\mathcal{T}_{GI}$ by the inversion of a tridiagonal matrix.

This treatment raises a key question. If $dQ/dt = 0$, how can the disk ever stabilize? In the course of solving $f_\mathrm{transport} = -f_\mathrm{source}$, sometimes a non-physical value of $\mathcal{T}$ will be obtained. In particular, since viscous heating $\propto -\mathcal{T}$ for reasonable rotation curves $\beta<1$, it must be the case that $\mathcal{T}\le 0$ to satisfy the second law of thermodynamics (turbulence should not decay into large-scale coherent motions). If this condition is not satisfied by the solution of $f_\mathrm{transport} = -f_\mathrm{source}$, then we set $\mathcal{T} = 0$ in that cell. Under this circumstance the cell behaves exactly as if it has stabilized, and $Q$ in that cell will obey $dQ/dt = f_\mathrm{source}$. Typically the reason that a cell falls into this situation is that $f_\mathrm{source} > 0$ and no physical value of $f_\mathrm{transport}$ can cancel this effect, so $Q$ is allowed to rise in that cell.

The gravitational stability of disks to linear axisymmetric perturbations is roughly determined by the value of $Q_\mathrm{Toomre}$. Modern versions of this parameter take into account both gas and stars \citep[e.g.][]{Rafikov2001,Romeo2013}, the finite thickness of the disk \citep{Shu1968,Romeo1992,Romeo1994,Elmegreen2011}, gas turbulence \citep{Hoffmann2012} and the fact that gas which can cool to arbitrarily small scales is never formally stable \citep{Elmegreen2011}. \citet{Romeo2011} have developed an approximate, but analytic, formula for $Q$ taking into account two components of finite thickness. To account for the final complication, we demarcate the stable from the unstable values of $Q$ at $Q_{GI}=2$, rather than the canonical value of unity, as suggested by \citet{Elmegreen2011}. This approximation to $Q$ and its partial derivatives with respect to $\Sigma$, $\sigma$, $\Sigma_*$, and $\sigma_*$ is extremely cheap to compute, which is advantageous since all of these values must be computed at each (unstable) radius and time to solve $f_\mathrm{transport}=-f_\mathrm{source}$.

Numerical experiments \citep{Stone1998, MacLow1998} of turbulent gas in periodic boxes have shown that the turbulence decays in roughly a crossing time of the turbulent driving scale. For the purposes of our simulations, we assume that the driving scale is the scale height of the disk, in which case the kinetic energy surface density $(3/2) \Sigma \sigma_{turb}^2$ will decay at a rate,
\begin{equation}
\mathcal{L} = \eta \Sigma\sigma^2\kappa Q_g^{-1}\left(1+ \frac{\sigma \Sigma_*}{\sigma_{zz}\Sigma}   \right) \left( 1 - \frac{\sigma_{th}^2}{\sigma^2}\right)^{3/2},
\end{equation}
where $\eta$ is a free parameter which would be $3/2$ if the decay time were exactly one scale height crossing time. As a result of the final factor, $\mathcal{L} \rightarrow 0$ as $\sigma \rightarrow \sigma_{th}$, i.e. when the gas ceases to be turbulent, it will no longer lose energy. The thermal velocity dispersion is set to correspond to a gas kinetic temperature of $7000 K$, the temperature of the warm neutral medium in the Milky Way. Physically, gas at this temperature will still radiate, but we assume that this radiation is balanced by energy injection. In the Milky Way's ISM, this is a balance between grain photoelectric heating and both Ly$\alpha$ and metal line cooling. In regions where the gas is largely molecular, this value of $\sigma_{th}$ no longer makes any sense, but these are also the regions where the disk is unstable because $\Sigma$ is large. In these regions the velocity dispersion $\sigma \gg \sigma_{th}(T=7000 K)$, so its value does not matter.

\subsection{Rotation curve}
\label{sec:rotCurve}

In order to derive the evolution equations shown in the previous section, we assumed that the potential and rotation curve of the disk are constant in time. The primary reason for this is that to self-consistently calculate $v_\phi$ would require knowledge of the dark matter. While N-body simulations assuming $\Lambda$CDM cosmology consistently produce dark matter halos with well-characterized density profiles, the effects of baryons are highly controversial. Moreover, if one were to calculate the rotation curve simply from the dark matter, \citep[e.g.][]{Cacciato2012}, the circular velocity would decrease with time since $z=1$ (at $z=2$, $v_{circ}\approx 185$ km s$^{-1}$, increasing to $\approx 200$ km s$^{-1}$ at $z=1$, and falling back to $\approx 190$ km s$^{-1}$ at $z=0$), whereas observations \citep{Kassin2012} show that (at fixed stellar mass) the circular velocity actually increases from $z=1$ to the present. Therefore rather than constructing a model for the rotation curve which depends on the poorly constrained interactions between baryons and dark matter, we adopt a simple functional form,
\begin{equation}
v_\phi(r) =v_{circ} \left( 1 + \left(r_b/r\right)^{|\beta_0 n |}\right)^{-\mathrm{sign}(\beta_0)/n}.
\end{equation}
This is designed to represent a smooth transition from powerlaw to flat, where $r_b$ is the characteristic radius where the rotation curve turns over. Within this radius, the velocity approaches a powerlaw with index $\beta_0$, and the sharpness of the transition between powerlaw and flat increases with increasing $n$. The disadvantages of this approach are that we are restricted to evolving our galaxies over periods during which the circular velocity does not change very much ($z\sim 2$ - 0), and changes to the potential owing to the movement of baryons are not reflected in the rotation curve.

\subsection{Star formation}
\label{sec:SF}
Stars form with a constant efficiency per freefall time $\epsilon_\mathrm{ff}$ from molecular gas, so that $\dot{\Sigma}_*^{SF} \sim \epsilon_\mathrm{ff} f_{{\rm H}_2} \Sigma / t_\mathrm{ff}$, where $t_\mathrm{ff}$ is the freefall time and $f_{{\rm H}_2}$ is the molecular fraction \citep{Krumholz2007,Krumholz2012}. Following \citet{Krumholz2012}, we posit that there are two regimes: one in which the appropriate time-scale is the freefall time of gas distributed over the full scale height $H$ of the disk, namely $t_\mathrm{ff} =  \sqrt{3\pi/32 G \rho} \approx \sqrt{3\pi H/ 32 G \Sigma}  $, which we call the `Toomre regime' and one in which the time-scale is determined by the freefall time of individual molecular clouds, which observations suggest is $t_\mathrm{ff} / \epsilon_\mathrm{ff} = \Sigma_{{\rm H}_2} / \dot{\Sigma}_*^{SF} \equiv t_{SC} \approx 2\ \mathrm{Gyr}$ \citep{Bigiel2011a}, the `single cloud regime'. Then the star formation rate is simply set by which of these two time-scales is shorter \footnote{Note that F12 omitted the factor of $1/Q_g\pi$, though the code and the appendix with the dimensionless version were correct},
\begin{equation}
\label{eq:sf}
\dot{\Sigma}_*^{SF} = \max\left(\epsilon_{\mathrm{ff}}f_{{\rm H}_2}\Sigma\kappa\frac{\sqrt{32/3}}{Q_g \pi}\left(1 + \frac{\sigma \Sigma_*}{\sigma_{zz}\Sigma}\right)^{1/2},\ \  f_{{\rm H}_2} \frac{\Sigma}{t_{SC}} \right)
\end{equation}
Typically the first regime is relevant at small radii since $\kappa \propto 1/r$, and the transition tends to be fairly constant in time, since the rotation curve is fixed in our model, and both terms are proportional to $f_{{\rm H}_2}$ and $\Sigma$, though $Q_g$ can change by an order of magnitude or more if the disk has stabilized.

The molecular fraction $f_{{\rm H}_2}$ is calculated according to the analytic formula of \citet{Krumholz2009c}. Their formula predicts $f_{{\rm H}_2}$ as a function of $\Sigma$ and $Z$. Roughly speaking, $f_{{\rm H}_2} \rightarrow 1$ at high surface densities, and below some transition surface density, there is a sharp cutoff where $f_{{\rm H}_2}$ rapidly approaches zero. This transition is metallicity dependent, roughly $5 M_{\odot}\ pc^{-2} (Z/Z_\odot)^{-1}$. We include the slight modifications to this formula we used in F12, namely a floor of $f_{{\rm H}_2} \ge 0.03$ to account for the fact that star formation is observed even at very low surface densities, \citep{Bigiel2010, Schruba2011}, likely as a result of the requirement that the FUV flux not fall below a certain floor in order for two-phase equilibrium in the atomic ISM to be possible \citep{Ostriker2010}.

At each time step, a new population of stars is formed with surface density $\dot{\Sigma}_*^{SF} dt$, where $dt$ is the duration of the time step. The velocity dispersion of this population is the maximum of $\sigma_{turb}$ and $\sigma_{*,min}$. Physically this floor might correspond to some combination of cloud-to-cloud velocity dispersion or the internal velocity dispersion of a cloud, roughly 2 km s$^{-1}$. The newly formed stars are then merged with the extant population while conserving mass and kinetic energy, meaning $\dot{\Sigma}_*^{SF} dt$ is added to $\Sigma_*$, and the velocity dispersion of the extant population is updated so that $(\Sigma_*\sigma_*^2)^\mathrm{new} = (\Sigma_*\sigma_*^2)^\mathrm{old} + \left(\dot{\Sigma}_*^{SF} dt\max \left( \sigma_{turb}^2, \sigma_{*,min}^2\right)\right)$. 

Once stars form, they also migrate. In our model, this is treated quite similarly to the gas migration discussed in section \ref{sec:gasTransport}, namely the stars experience torques if they are gravitationally unstable to spiral instabilities. Our prescription has improved significantly since F12, so we discuss the new governing equations in appendix \ref{app:stmig}. Overall this typically has a minor effect on the dynamics of the disk, although it can strongly influence the stellar velocity dispersions particularly at small radii.

\subsection{Metallicity}
\label{sec:metallicity}

In addition to its dynamical effects, star formation is responsible for the production of metals. We approximate this process as instantaneous, in which case the production of metals is proportional to the star formation rate. In each cell the mass in metals is evolved according to 
\begin{eqnarray}
\frac{\partial M_Z}{\partial t} &=& \Delta r \frac{\partial \dot{M} Z}{\partial r} + (y f_R  - f_R Z  - \mu Z_w ) \dot{M}_*^{SF} \nonumber \\
& &  + \dot{M}_{acc} Z_{IGM} + \frac\partial{\partial r} \kappa_Z \frac\partial{\partial r} M_Z.
\end{eqnarray}
The first term accounts for metals advected from other parts of the disk; $\Delta r$ is defined as the width of the cell under consideration. The next term includes three effects which occur in proportion to the star formation rate in that cell, $\dot{M}_*^{SF} \equiv \pi (r_{i+1/2}^2 - r_{i-1/2}^2) \dot{\Sigma}_*^{SF}$  - here $r_{i+1/2}\equiv\sqrt{r_ir_{i+1}}$, the location of the boundary between cells $i$ and $i+1$ on our logarithmic grid. The first is the production of new metals through the course of stellar evolution, which occurs in proportion to $y$, defined as the mass of metals produced per unit mass (of all gas) locked in stars. Next is the mass of metals locked in stellar remnants. The final term proportional to the star formation rate is the mass of metals ejected in galactic winds with mass loading factor $\mu$.  Defining $\dot{M}_{acc} \equiv\pi(r_{i+1/2}^2-r_{i-1/2}^2) \dot{\Sigma}_{cos} $, the next term is simply the mass of metals accreting from the IGM. The final term, metal diffusion, will be discussed momentarily.

The metallicity of the wind is given by $Z_w$. Many authors assume that $Z_w=Z$, the metallicity of the gas in the disk. It is worth pointing out that this is probably a lower bound, but there is also an upper bound. In the limit of small mass loading factor $\mu$, the maximum metallicity is the mass in metals expelled by stellar winds and supernovae: $(y f_R + (1-f_R)Z)\Delta M_*$ divided by the total mass ejected, $(1-f_R) \Delta M_*$.  When the mass loading factor is larger than $1-f_R$, some additional mass from the ISM must also be swept up, thereby decreasing the maximum metallicity. The metallicity of the wind must therefore be
\begin{equation}
Z < Z_w < \left\{ \begin{array}{cc}
	 Z + y f_R/(1-f_R) & \mbox{ if $\mu \le 1-f_R$} \\
	 Z + y f_R/\mu & \mbox{ if $\mu > 1-f_R$}
	 \end{array} \right.
\end{equation}
We therefore define a new parameter $\xi$, similar in spirit to e.g. the metal loss factor in \citet{Krumholz2012b}, so that
\begin{equation}
Z_w = Z + \xi \frac{y f_R}{ \max(\mu, 1-f_R)}.
\end{equation}
Here $\xi$ may vary between 0 and 1, with 0 representing the usual assumption of perfect mixing of stellar ejecta and galactic outflows, and 1 representing the minimal possible mixing.

The diffusion of metals has received relatively little attention until recently. In F12, we included this diffusion term to prevent the metallicity gradient from steepening excessively, tuning the value of $\kappa_Z$ to yield a reasonable gradient. Since then, \citet{Yang2012} have measured the value of $\kappa_Z$ in a 2D shearing box simulation with turbulence driven by thermal instability. They show that to a reasonable approximation $\kappa_Z \propto r_{inj}^2 / t_{orb}$, where $r_{inj}$ is the initial wavelength of the metallicity perturbation, and $t_{orb}$ is the orbital time. Here we make the approximation that $r_{inj} \approx \lambda_J = \sigma^2/G\Sigma$, the 2D Jeans length, since this should be similar to the spacing of the largest giant molecular clouds. We can therefore scale $\kappa_Z$ in our simulation to their measured value, as
\begin{equation}
\label{eq:Zdiff}
\kappa_Z(r,t) = k_Z 1.2\ \frac{\mathrm{kpc}^2}{\mathrm{Gyr}} \left(\frac{\sigma^2 / G \Sigma}{3.1\ \mathrm{kpc}}\right)^2 \frac{\kappa}{\sqrt{2}\ (26\ \mathrm{km\ s^{-1} kpc^{-1}})}
\end{equation}
The numerical values are the measured $\kappa_Z$ and the input parameters $r_{inj}$ and $\Omega$ quoted for one of their simulations. We also include a free parameter $k_Z\approx 1$, recognizing that there is some uncertainty in this result. The numerical implementation of the diffusion term is operator-split from the rest of the terms, implicit, and computed in terms of fluxes so that metal mass is explicitly conserved. We also enforce $\kappa_Z < v_{circ} R$, the largest velocity and radius in the problem, which is not guaranteed by equation \ref{eq:Zdiff} when $\Sigma$ is very small. This essentially makes sure that the metal injection scale $r_{inj} \la R$, the size of the system.

\subsection{Accretion}
\label{sec:acc}
In our model, gas accretes onto the disk at an externally-prescribed rate $\dot{M}_{ext}$ and a profile $\dot{\Sigma}_{cos}$ such that 
\begin{equation}
\dot{M}_{ext}(t) = \int_0^{\infty} \dot{\Sigma}_{cos}(r,t) 2\pi r dr .
\end{equation}
In our fiducial model we take $\dot{\Sigma}_{cos} \propto \exp(-r/r_{acc}(z))$. The angular momentum of accreting gas is thereby entirely set by $r_{acc}(z)$, which is assumed to scale with halo mass so that 
\begin{equation}
r_{acc}(z) = r_{acc}(z=0) \left(\frac{M_h(z)}{M_h(z=0)}  \right)^{\alpha_r},
\end{equation}
with $r_{acc}(z=0)$ and $\alpha_r$ left as free parameters. A reasonable guess for $\alpha_r$ is $1/3$, which roughly corresponds to the assumption that $r_{acc} \propto R_{vir}$ \citep[e.g.][]{Mo1998}, while a reasonable guess for $r_{acc}$ might be the size scale of local disk galaxies, which varies significantly at fixed mass but is of order 10 kpc. 

To determine $\dot{M}_{ext}$ at each time step in our simulation, we calculate $M_h(t)$, the history of the dark matter halo mass, differentiate with respect to time, and multiply by $f_b \epsilon_{in}(M_h,z)$, where $f_b\approx 0.17$ is the universal baryon fraction, and $\epsilon_{in}$ is some efficiency. We take two separate approaches to calculating $M_h(t)$. The first is to use an average dark-matter accretion history \citep{Neistein2008a}, which estimates the average growth rate to be
\begin{equation}
\label{eq:avgAcc}
\dot{M_h} = 39 (M_h / 10^{12} M_\odot)^{1.1} (1+z)^{2.2} M_\odot / yr,
\end{equation}
which agrees well with hydrodynamic simulations \citep{Dekel2013}. This approach allows us to quickly and clearly see the effects of changes in the physical parameters of the simulations without averaging over many galaxies with different accretion histories. The disadvantage is that in reality galaxies are likely to have stochastic accretion histories, and this will have a significant effect on the resultant galaxies. For instance, if a galaxy is fed at a steady rate, if a given region of the disk becomes stable to gravitational turbulence it is unlikely to ever destabilize again, but an accretion history with variation about the median could be unstable at low redshifts or stable at high redshifts.

To capture the effects of variable accretion histories, we also generate accretion histories using the analytical EPS-like formalism developed by \citet{Neistein2008a} and \citet{Neistein2010}. The procedure is as follows. The desired final halo mass $M_{h,0}$ and redshift ($z=0$) are converted into their corresponding dimensionless values $S$ and $\omega$. We use the approximate relation from \citet{VandenBosch2002}
\begin{equation}
\label{eq:SMh}
S(M_h) = u^2 \left( \frac{c_0 \Gamma}{\Omega_m^{1/3}} \left(\frac{M_h}{1 M_\odot}\right)^{1/3}  \right) \frac{\sigma_8^2}{u^2(32\Gamma)}.
\end{equation}
The parameters $c_0$ and $\Gamma$ are respectively $3.804\times 10^{-4}$ and $0.169$. The function $u(x)$ is given by
\begin{eqnarray}
u(x) &=& 64.087 \left( 1 + 1.074\ x^{0.3} - 1.581\ x^{0.4} \right.  \nonumber \\
& & \left. + 0.954\ x^{0.5} - 0.185\ x^{0.6}  \right)^{-10}.
\end{eqnarray}
Meanwhile, $\omega(z)$ may be computed approximately \citep{Neistein2008a} by 
\begin{equation}
\label{eq:omega}
\omega(z) = 1.260 \left( 1 + z + 0.09(1+z)^{-1} + 0.24 e^{-1.16 z} \right).
\end{equation}
With these relations, we now have $S(M_h(z=0))$, and $\omega(z=0)$.

The independent variable $\omega$ is steadily incremented by a fixed value $\Delta \omega$ until the entire desired redshift range is encompassed. At each step in $\omega$, a new value of S is computed by adding
\begin{equation}
\label{eq:accLognormal}
\Delta S = \exp\left(x \sigma_k + \mu_k\right)
\end{equation}
where $x$ is a value drawn from a normal distribution with zero mean and unity variance. We use a fixed $\Delta\omega=0.1$, since this is the timestep used in generating the fitting formulae for $\sigma_k$ and $\mu_k$ in \citet{Neistein2008a}. The fact that we use a fixed $\Delta\omega$ rather than a distribution leads to the distinct steps in Fig. \ref{fig:accHistory}, where all of the accretion histories change at once.

The mean of the normal distribution to be exponentiated, $\mu_k$, and its standard deviation $\sigma_k$, depend on halo mass, and are fit to the results of the Millenium Run \citep{Springel2005}.
\begin{eqnarray}
\label{eq:sigk}
\sigma_k &=& 1.367+0.012\log_{10}S +0.234(\log_{10}S)^2  \\
\label{eq:muk}
\mu_k &=&  -3.682+0.76\log_{10}S - 0.36(\log_{10}S)^2
\end{eqnarray}
Converting each value of $S$ back to $M_h$ one obtains a dark matter accretion history $M_h(\omega_j)$, where the $\omega_j$ are the sequence of $\omega$'s obtained by incrementing $\omega$ by the fixed $\Delta \omega$, namely $\omega_j = \omega_0 + j\Delta\omega$ for $j=0,1,2,...$ and $\omega_0 = \omega(z=0)$. We require that the change in $M_h$ over a single step, $M_h(\omega_i)-M_h(\omega_{i+1})$, not exceed $M_h(\omega_{i+1})$ to avoid galaxies `accreting' a larger mass than their own, i.e. becoming a satellite. Since equations \ref{eq:sigk} and \ref{eq:muk} were obtained by a fit to the Millenium Run using a cosmology where $(\Omega_m, \sigma_8) = (0.25,0.9)$, when converting between $M_h$ and $S$ with equation \ref{eq:SMh}, we use the parameters from the Millenium Run. Once we have obtained $M_h(\omega_j)$ using this cosmology, we can transform it so that it agrees with the WMAP5 \citep{Komatsu2009} cosmology $(\Omega_m,\sigma_8)=(0.258,0.796)$, which is much closer to the current best-fitting values. We use the scaling obtained in \citet{Neistein2010} via a comparison of merger trees from Millenium and an N-body simulation run with WMAP5 cosmology, namely we replace the $\omega_j$ with $\tilde{\omega}_j = \omega_0 + 0.86 j \Delta \omega$. The full dark matter mass history of the halo $M_h(t)$ is then obtained by converting the $\tilde{\omega}_j$ to $z$ (with equation \ref{eq:omega}) and subsequently to $t$, and linearly interpolating the sequence of halo masses $M_h(\tilde{\omega}_j)$ in time. The dark matter accretion history is then just the instantaneous derivative of $M_h(t)$.
 
\begin{figure}
 	\centering
	\includegraphics[width=8 cm]{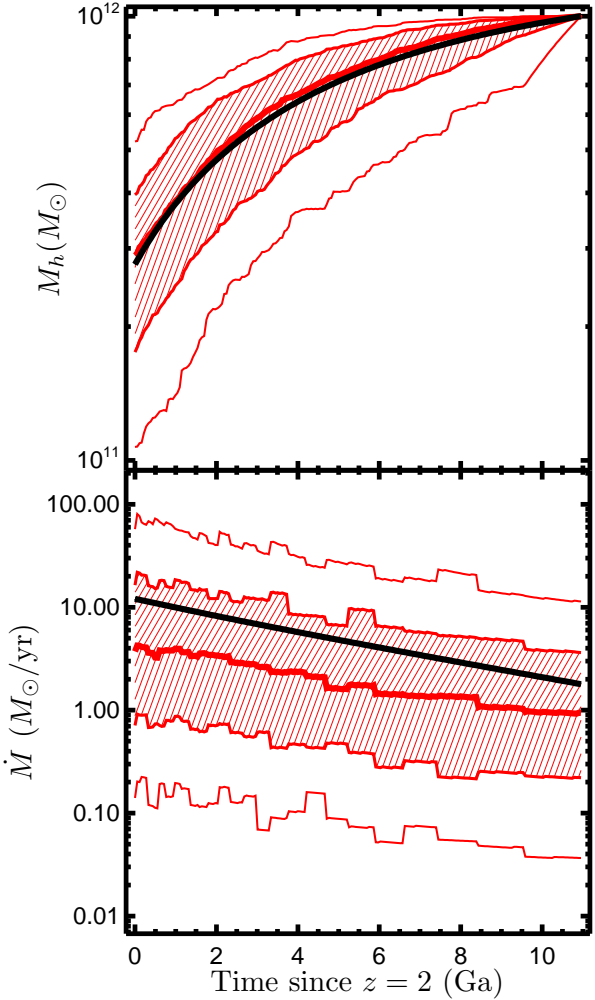} 
	\caption{The growth of halos. The top panel shows the evolution of the halo mass for the smooth accretion history (black) and the median, central 68 per cent (shaded), and central 95 per cent of 400 stochastic accretion histories (red). The corresponding distribution of the inferred baryon accretion rates, including the efficiency factor (equation \ref{eq:accEff}), is shown in the bottom panel. The steps in $\dot{M}$ correspond to our fixed interval $\Delta\omega=0.086$.}
	\label{fig:accHistory}
\end{figure}
 
The input $\dot{M}_{ext}(t)$ to our simulation is taken to be the average dark matter accretion rate at time t, generated either from the smooth accretion formula \ref{eq:avgAcc} or the lognormal one \ref{eq:accLognormal} times $f_b\epsilon_{in}$. For the efficiency, we use a reasonably general parameterization,
 \begin{equation}
 \label{eq:accEff}
 \epsilon_{in}(M_h,z) = \min\left(\epsilon_{0} \left(\frac{M_h}{10^{12}} \right)^{\beta_{M_h}} (1+z)^{\beta_{z}},\ \ \epsilon_{max} \right)
 \end{equation}
 \citet{Faucher-Giguere2011} fit the results of a cosmological SPH simulation with no feedback to find $(\epsilon_0,\beta_{M_h},\beta_z,\epsilon_{max})=(0.31, -0.25, 0.38,1)$, though they explicitly only use this fit $\it above$ $z=2$.  
 
Despite its success at high redshift, the paradigm of cold accretion is fairly uncertain for galaxies which have some hot coronal gas, like the Milky Way, at low redshift. Moreover, \citet{Diemer2013} have pointed out that below $z\sim 1$, nearly all the growth in $M_h$ for haloes with $M_h(z=0)\sim 10^{12} M_\odot$ corresponds to the fact that the background density of the universe is decreasing (roughly as $\rho_m \propto (1+z)^3$) while dark matter halos are changing very little. Because the halos are defined in simulations as having a spherical overdensity relative to the background of $\sim 200$, relatively static halos increase their mass merely because of this drop in the background density. \citet{Dekel2013} have verified that this is not a significant effect at $z>1$.
 
A number of ideas have been proposed to explain how MW-like galaxies can maintain star formation rates of order $2 M_\odot\ yr^{-1}$ despite little evidence of cold accretion at anything near these rates. The gas may be accreting in an ionized phase, slightly hotter than the observed High Velocity Clouds in HI \citep{Joung2012a}. The process may be helped along by supernova-induced accretion, where hot halo gas is supposed to condense in the wakes of cold clouds ejected by supernova feedback from the disk of the galaxy \citep{Marinacci2010}. Alternatively galaxies can be powered by gas recycled back to the ISM from stars \citep{Leitner2011}; while much of this process can be approximated as occurring instantaneously (the winds from and supernovae of massive stars), a significant amount of mass is returned even from very old stellar populations \citep[see also][]{Martig2010}. Gas ejected by galactic winds often finds its way back to the star forming disk \citep{Oppenheimer2010}, which may provide yet another way to provide star-forming gas to galaxies even if dark matter is not accreting.

Given the uncertainties in how gas is accreted at low redshift, our naive approach of setting $\dot{M}_{ext} = \dot{M}_h f_b \epsilon_{in}$ is not unreasonable. In our fiducial model, a MW-mass galaxy accretes roughly $2 M_\odot\ yr^{-1}$ at redshift zero, and so yields a star formation rate similar to observations, even if the physical mechanism for this accretion is unclear. We do retain, in varying the parameters of the accretion efficiency and the accretion profile, a considerable amount of flexibility in the model, which is appropriate given the uncertainties.  In Fig. \ref{fig:accHistory}, we show $M_h(t)$ and the resulting $\dot{M}_{ext}$ for the fiducial smooth model and the stochastic accretion model.

\subsection{Initial conditions at $z \sim 2$}
\label{sec:IC}
Having constructed the accretion history, we can now generate an initial condition. To do so, we first require that the total surface density in gas and stars equals some fraction $f_{cool}$ of the total baryonic mass available, $f_b M_h(z=z_{start})$. For haloes which will host a single galaxy at redshift zero, it is reasonable to assume that at high redshift, $M_h(z=z_{start})$ will be small enough that the cooling time of halo gas is short, and that even if a galaxy has a stable virial shock, it may still be fed by cold streams, and so $f_{cool}$ should be of order unity \citep{Birnboim2003,Keres2005,Dekel2006,Ocvirk2008,Dekel2009,Danovich2012,Dekel2013}.

We next make the fairly arbitrary decision to have a fixed initial gas fraction $f_{g,0}$, defined at each radius to be $f_g(r) = \Sigma/(\Sigma+\Sigma_*)$. Thus $\Sigma$ and $\Sigma_*$ will have the same shape. Observations of main sequence galaxies at high redshift show their stellar profiles to be exponential \citep{Wuyts2011a,vanDokkum2013}, so we choose an initial exponential profile with scale length $r_{IC} = r_{acc}(z=z_{start})$. With these requirements we arrive at the initial profile,
\begin{equation}
\Sigma =  f_{g,0} f_{cool} f_b \frac{M_h(z=z_{start})}{2\pi r_{IC}^2} \exp\left(-r/r_{IC}\right)\frac{1}{1-f_{out}}
\end{equation}
The final factor is a correction for the finite size of the computational domain. In particular, we want the initial mass of the disk to be independent of $R$, so $f_{out}$ is the fraction of the mass profile which lies beyond the computational domain,
\begin{equation}
f_{out} = \frac{1}{2\pi r_{IC}^2} \int_R^\infty 2\pi r\ e^{-r/r_{IC}} dr 
\end{equation}
Since the initial conditions are highly uncertain, it is more important to get the correct amount of mass in the computational domain than to make sure the profile has a particular normalization. Still, we typically set $R \gg r_{IC}$ so that this is a minor correction.

The other initial variables we need to specify are $\sigma$, $\sigma_{rr}$, $\sigma_{zz}$, $Z$, and $Z_*$. For the metallicities, we simply set $Z=Z_*=Z_{IGM}$. For the velocity dispersions, we use $\sigma=\sigma_{rr}\phi_0^{-1}=\sigma_{zz}\phi_0^{-1}=\sigma_{th}$, i.e. the value of our minimum velocity dispersion. We allow the velocity dispersion of the stars to be different (generally higher) than that of the gas, with a free parameter $\phi_0$. The low constant values of the velocity dispersion will often lead some parts of the disk to have $Q<Q_{GI}$, so in those regions we raise $\sigma$, $\sigma_{rr}\phi_0^{-1}$ and $\sigma_{zz}\phi_0^{-1}$ simultaneously (keeping them equal) until $Q=Q_{GI}$. We emphasize, though, that the gas velocity dispersion $\sigma$ and the two stellar velocity dispersions $\sigma_{rr}$ and $\sigma_{zz}$ evolve separately throughout the simulation - their ratio is fixed only initially. The idea is that, since supersonic turbulence in the disk is generated exclusively by gravitational instability in our model, any region not subject to this instability will have $\sigma\approx \sigma_{th}$. 

Typically our initial conditions have $Q=Q_{GI}$ in some annulus. At larger radii $\Sigma$ drops off quickly so $Q \propto \Sigma^{-1}$ increases, while $Q$ also increases at smaller radii through the dependence $\kappa \propto v_\phi/r$. We discuss the (lack of) sensitivity of our results to these choices of initial conditions in appendix \ref{app:vary}.

\section{Simulation Results}
\label{sec:results}
In this section we discuss some generic features of the galaxies produced by our model. We begin by exploring models with smooth accretion histories and a fiducial choice of parameters, which we summarize in Table \ref{ta:params}. These are compared with artificial, illustrative models where one important physical ingredient is turned off by hand. We then allow the accretion histories to vary stochastically in a cosmologically realistic way, illustrating the differences between galaxies with identical physical laws but different accretion histories, as one might expect for real galaxies. Finally we compare our models with recent observational results.

\subsection{Equilibria in smoothly accreting models}
There are three terms in the continuity equation (equation \ref{eq:dcoldt}). At a particular radius, gas arrives via $\dot{\Sigma}_{cos}$, departs via $(f_R+\mu)\dot{\Sigma}^{SF}$, and moves to or from other radii via $\dot{\Sigma}_\mathrm{tr} \equiv (2\pi r)^{-1} (\partial \dot{M} / \partial r)$. The generic behavior of this equation at a given radius in our fiducial model is that gas will build up, either via direct accretion or as mass arrives from somewhere else in the disk, until an equilibrium is reached such that $\dot{\Sigma} \approx 0$. This equilibrium will then slowly evolve with time as the global gas accretion rate $\dot{M}_{ext}$ falls off. 

\begin{figure*}
	\centering
	\includegraphics[width=17 cm]{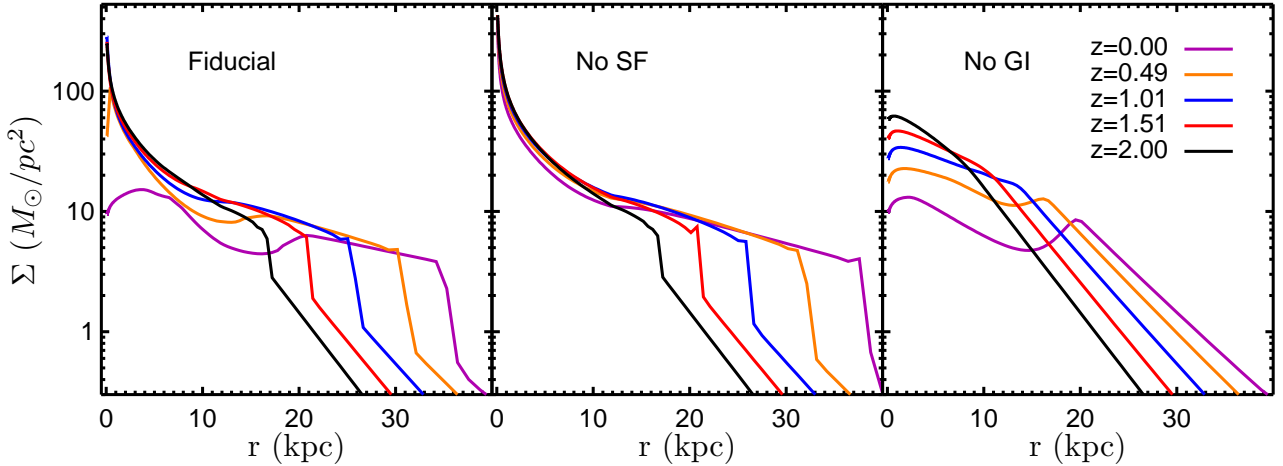} 
	\caption{The three simplified models. For each of the these simplified models, we show the evolution of the radial gas surface density distribution. We see that the evolution of the fiducial model is a non-trivial combination of the effects of star formation and GI transport. }
	\label{fig:simpleCol}
\end{figure*}

To aid in understanding how this equilibrium emerges, we have run three simple models with identical smooth accretion histories - (i.) the fiducial model - our best guess for physical parameters which will lead to something resembling the Milky Way (see Table \ref{ta:params}), (ii.) the same model with no star formation, and (iii.) the same model with no gravitational instability, i.e., $\mathcal{T}_{GI} = 0$ everywhere. The features of models i. and ii. are similar at large radii, while the features of i. and iii. bear some resemblance at small radii. This immediately suggests that GI transport is important at large radii and star formation is important at small radii. The gas  surface density distributions of each model are shown in Fig. \ref{fig:simpleCol} as a function of time. The gas is supplied via an exponential distribution, $\dot{\Sigma}_{cos} \propto e^{-r/r_{acc}}$. Without gravitational instability (model iii), star formation carves out the inner parts of the distribution, leaving a hole in the gas at galactic centers, while without star formation (ii), gas is redistributed into a powerlaw distribution, following roughly $\Sigma \propto 1/r$.

\begin{figure*}
	\centering
	\includegraphics[height=17 cm]{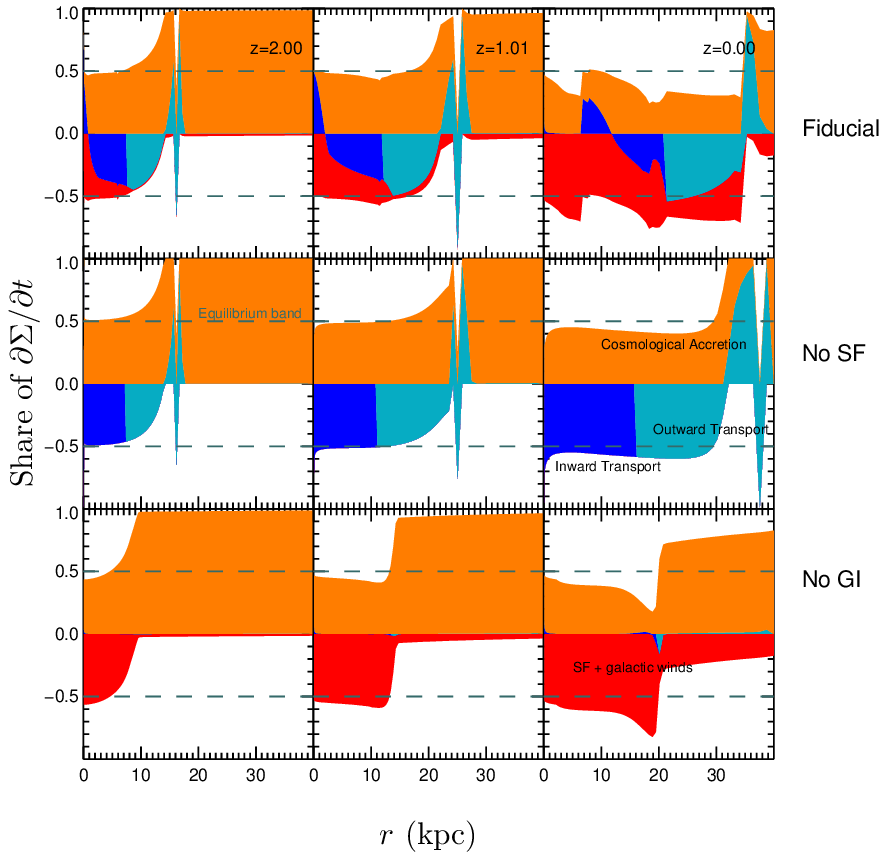} 
	\caption{The balance of terms in the continuity equation. The terms contributing to $\partial \Sigma/\partial t$ are split into those which increase $\Sigma$ at a particular radius and time and those which decrease $\Sigma$. The former appear above zero and the latter below. Each term is represented by a different color - orange for $\dot{\Sigma}_{cos}$, red for $(\mu+f_R)\dot{\Sigma}^{SF}_*$, and blue for $\dot{\Sigma}_\mathrm{trans}$. Light (dark) blue indicates gas being transported outwards (inwards). At each radius the height of the colored band is normalized to unity, and its position shows how close the disk is to equilibrium (equal positive and negative contributions) at that radius. The columns show different redshifts ($z=2,1,0$) and the rows show different models: fiducial, no star formation, and no GI transport. The feature at e.g. $r=15$ kpc and $z=2$ comes from gas from the unstable region heating when it piles up in a single stable cell. Real galaxies will not have such a sharp transition since there will be some breaking of axisymmetry and some overshoot from the unstable region, neither of which we model here. The feature at $z=0$ around 15-20 kpc in the fiducial model is caused by a feature in the surface density distribution - this section of the galaxy has had the direction of GI transport reverse from outward to inward.  }
	\label{fig:simpleBal}
\end{figure*}

A useful way to understand what sets the surface density is to examine the relative effects of each term in the continuity equation. In particular, at each time and radius, we can divide each term by $A \equiv |\dot{\Sigma}_\mathrm{tr}| + \dot{\Sigma}_{cos} + (\mu + f_R)\dot{\Sigma}^{SF}$. In Fig. \ref{fig:simpleBal} we compute these contributions, including the sign of their effect on the overall value of $\partial \Sigma/\partial t$, so at each radius the fraction of the colored region occupied by (red, orange, blue) represents the fraction of $A$ from (star formation, cosmological accretion, transport). The different shades of blue show which way the mass is flowing in the disk, i.e. the sign of $\dot{M}$ - dark blue indicates gas flowing towards the center of the disk, and light blue outward motion.

When the colored band in Fig. \ref{fig:simpleBal} stretches from -0.5 to 0.5, that region of the disk has reached an equilibrium configuration. In each case shown here, the equilibration proceeds from inside outwards. This is a combination of two effects- the especially efficient star formation in the center of the disk, and the fairly centrally-concentrated distribution of accreting gas. The equilibrium does not last forever-  at $z=0$, there can be significant deviations as the disk processes past accretion and the instantaneous accretion rate falls owing to the expansion of the universe on time-scales potentially shorter than the gas depletion time at these large radii.

\subsubsection{Equilibrium between SF and accretion: the No GI Model}

We first focus on the model with no GI. In this model, at a given radius, gas builds up until the local star formation rate $\propto \Sigma$ can balance the incoming accretion. This happens first in the center of the disk. Not only is the cosmological accretion rate per unit area larger there, but the star formation time-scale is shortest (Fig. \ref{fig:tdep}). In this model there is in fact a huge range of depletion times, from roughly 100 Myr at $z=2$ at small radii to 60 Gyr in the outer disk. There are two effects driving this diversity. For depletion times between 100 Myr and 2 Gyr, the disk is in the Toomre regime of star formation (see equation \ref{eq:sf}), for which the depletion time scales as $\kappa^{-1}$. This region is typically small, $\la 3 $ kpc, outside of which the time-scale would become longer than $2 $ Gyr if it continued to follow the $\kappa^{-1}$ scaling. At this point the disk transitions to the single-cloud regime of star formation.  At the transition, the disk still tends to be dominated by molecular gas. In the mostly-molecular but still single-cloud regime, the depletion time is roughly 2 Gyr, the single-cloud molecular depletion time - this can be seen as a flattening in the $t_{dep}$ distribution with radius. There is then a transition from molecular to atomic gas, which accounts for the difference between parts of the disk with a 2 Gyr depletion time and a 60 Gyr depletion time - this maximum depletion time is set by $f_{{\rm H}_2,min}$, which is quite uncertain.

A generic feature of the No GI model is that at the edge of the star-forming region, star formation occurs at a slightly faster rate than new gas is accreted at that radius (Fig. \ref{fig:simpleBal}, bottom right panel). All of the models, particularly at lower redshift exhibit a slight tendency to fall just below the `equilibrium band' after they have initially equilibrated at a given radius, since the accretion rate is externally imposed and falling monotonically. The feature at $z=0$ in the No GI model goes beyond this, however, and may be explained by a small feedback loop in the star formation law introduced by the dependence on metallicity. The demarcation between the star-forming part of the disk and the outskirts is set by the molecular to atomic transition. Typically the star formation rate at a given radius is able to balance the incoming material only if the molecular fraction there is above the minimum allowed value - otherwise star formation would be too slow. When enough gas has accumulated to satisfy $f_{{\rm H}_2}>f_{{\rm H}_2,min}=0.03$, the star formation rate rises steeply with column density and new metals are produced, which in turn catalyze star formation by reducing the amount of gas needed to maintain a molecular, star-forming phase. Thus the extra gas, which is now no longer necessary for the star formation rate to balance the accretion rate, can be consumed, though this generally takes a significant amount of time, $t_{dep} \ga 2$ Gyr. 

\begin{figure}
	\centering
	\includegraphics[width=8 cm]{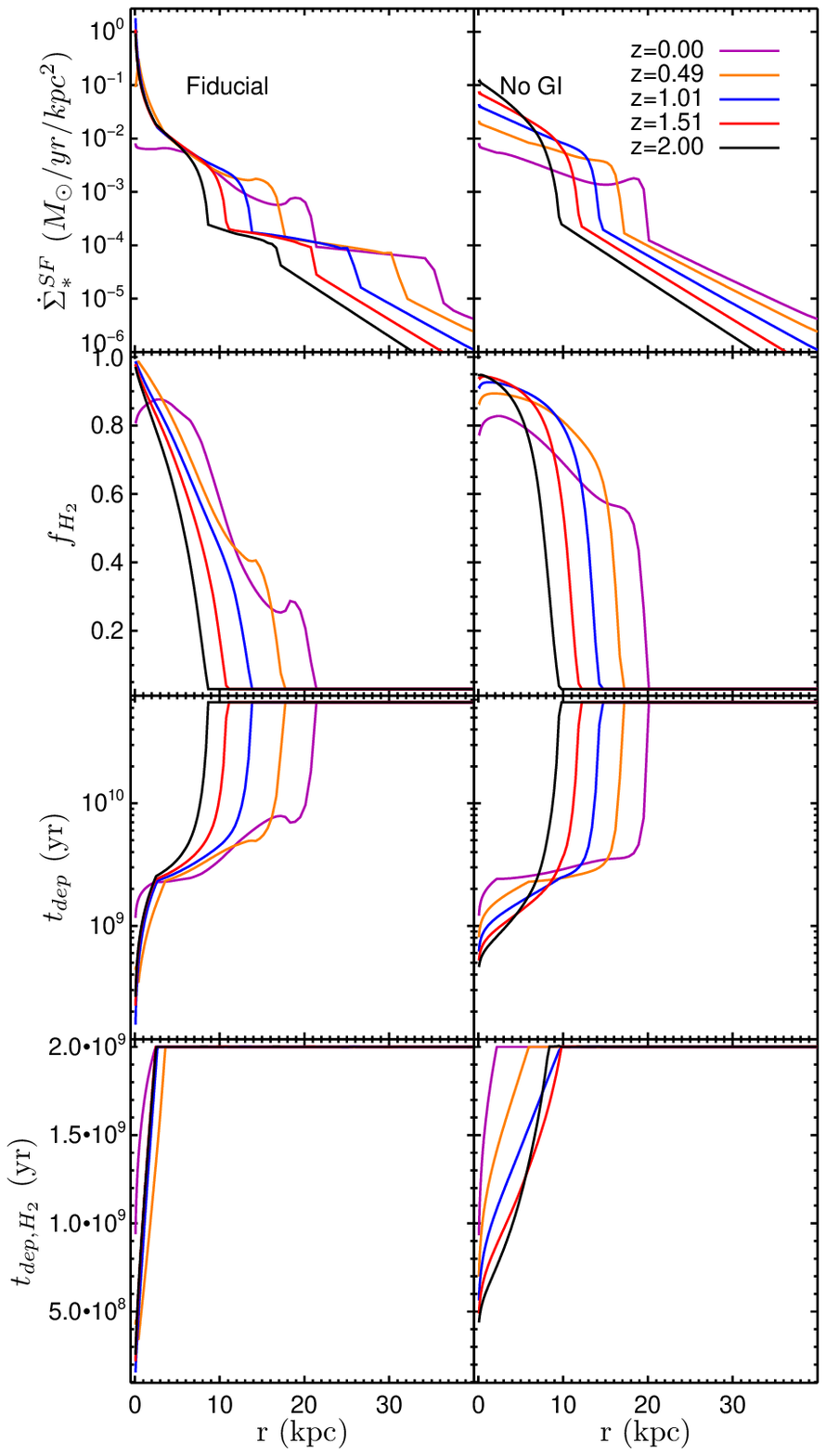} 
	\caption{Star formation in the fiducial and No GI models. The star formation rate (top) is proportional to the surface density of gas modulated by other factors reflected in the depletion time, $t_{dep} = \Sigma/\dot{\Sigma}^{SF}_*$ (second row). The first is the molecular fraction (third row), itself determined by $\Sigma$ and $Z$, and the second is the regime of star formation (bottom row), either single-cloud ($t_{dep,{\rm H}_2} = 2$ Gyr), or Toomre ($t_{dep,{\rm H}_2} \propto 1/\sqrt{G\rho} < 2$ Gyr). For each quantity, the left panel shows the fiducial model, while the right shows the model with GI transport turned off. Star formation in the fiducial model is much more concentrated and reaches much higher surface densities $\dot{\Sigma}^{SF}_*$ through the action of GI transport. The absence of GI causes so much gas to build up at larger radii that at high redshift the Toomre regime of star formation extends to nearly 10 kpc, instead of just the inner few kpc.}
	\label{fig:tdep}
\end{figure}

\subsubsection{Equilibrium between GI transport and accretion: the No SF Model}

We now turn to the no SF model to help us understand the importance of GI. In our model, when the disk has enough gas to be gravitationally unstable, it self-regulates to a marginally stable level, namely $Q=Q_{GI}=const.$, where $Q_{GI}$ demarcates gravitational stability from instability. The value of $Q$ depends on the surface densities and velocity dispersions of the gas and stars. In our numerical simulations we account for these dependences using the formula from \citet{Romeo2011}, but this formula reduces to something quite similar to the much simpler \citet{Wang1994} approximation when $\sigma\approx \sigma_{rr} \approx \sigma_{zz}$, namely $Q^{-1} \sim (2/3) (Q_g^{-1} + Q_*^{-1})$. In our model the situation can be simplified even further by the fact that $Q_*$ is separately self-regulated by stellar migration via transient spiral heating, so that $Q \sim (3/2) Q_g$. In this case the $Q=Q_{GI}$ condition may be re-written
\begin{equation}
\label{eq:sigmaGI}
\Sigma \approx \Sigma_{GI} \equiv \frac32 \frac{\sqrt{2(\beta+1)} v_\phi \sigma}{\pi G r Q_{GI}}
\end{equation}
At a given radius, $\beta$, $v_\phi$, $r$, and $Q_{GI}$ are all fixed, so equation \ref{eq:sigmaGI} may be considered a direct mapping between $\Sigma$ and $\sigma$. If $\sigma$ does not vary by much and the velocity dispersions of the gas and stars are similar, then $\Sigma$ will simply follow a $1/r$ powerlaw over a wide range of radii. 

The velocity dispersion and hence $r \Sigma_{GI}$ is restricted to a relatively narrow range because there is both a minimum and maximum velocity dispersion. The minimum is set by the thermal velocity dispersion, $\sigma_{th}$ -- the gas cannot get colder than when its turbulent velocity dispersion is zero. We can therefore say that in a gravitationally unstable region,
\begin{equation}
\Sigma \ga \Sigma_{crit} \equiv \frac32 \frac{\sqrt{2(\beta+1)} v_\phi \sigma_{th}}{\pi G r Q_{GI}}.
\end{equation}
The maximum is determined by the gas supply -- for a given $\dot{M}_{ext}$ to be transported to the center of the disk in a quasi-steady state, it must dissipate the gravitational potential energy between where it arrives and the center of the galaxy, and it must experience enough torque to lose its angular momentum. In a steady state, local heating by torques balances local cooling by turbulent dissipation (see section \ref{sec:energyEq}). Note that `heating' and `cooling' refer to changing the turbulent velocity dispersion of the gas, not its kinetic temperature. The rate at which the gas cools (and hence experiences torques) $\mathcal{L}$, depends on the velocity dispersion. The maximum velocity dispersion is therefore set by assuming that 100 per cent of the gas arriving from an external source flows towards the center in steady state. Since some gas never reaches the center because of star formation, and other gas moves outwards rather than inwards, this is an upper limit. As shown in section \ref{sec:maxSigma}, at $z \sim 2$ for galaxies accreting at $\sim 10 M_\odot\ yr^{-1}$ the velocity dispersion is restricted to 8 km s$^{-1}< \sigma \la  20$ km\ s$^{-1}$. This value is low compared to the measured velocity dispersions in the SINS galaxies. As we will see in section \ref{sec:maxSigma}, some small fraction of MW-progenitors do have much higher accretion rates in our stochastic accretion model. Moreover, the SINS galaxies are likely somewhat more massive than the MW progenitors we consider here.

 \begin{figure*}
	\centering
	\includegraphics[width=17 cm]{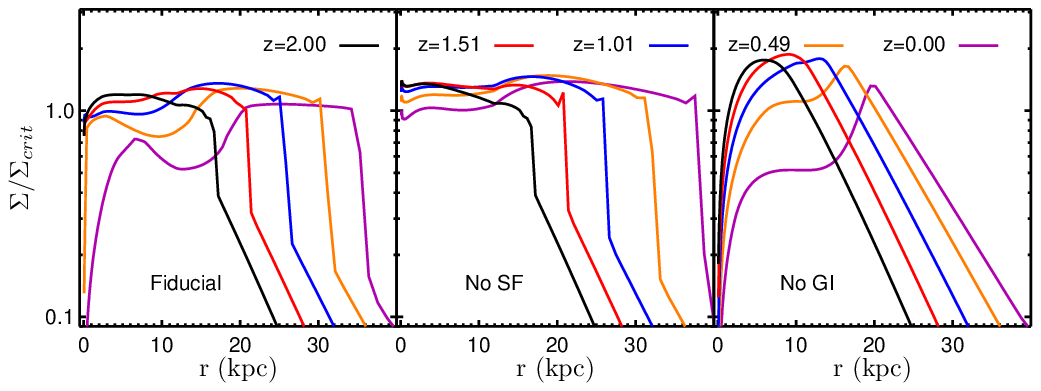} 
	\caption{The ratio of $\Sigma$ to the minimum surface density necessary for gravitational instability, $\Sigma_{crit}$. Gas spreads out to keep the surface density above this critical value but below the maximum value, $\Sigma_{GI}$ evaluated at $\sigma_{max}$. This ratio falls below $\sim 1$ where the disk has not yet destabilized (left two panels, large radii) or has stabilized due to GI quenching (left panel, small radii, low redshift). Note however that in the fiducial model, especially at low redshift, $\Sigma$ can fall slightly below $\Sigma_{crit}$ even in gravitationally unstable regions because in deriving $\Sigma_{GI}$ and $\Sigma_{crit}$ we assumed that $\sigma \approx \sigma_{rr} \approx \sigma_{zz}$, which is no longer true at low redshift. This is a factor of two level effect -- the gravitationally stable regions always have $\Sigma$ well below $\Sigma_{crit}$. Interestingly, even the `No GI' run does not reach values far larger than $\Sigma/\Sigma_{crit} = 1$ (although it reaches significantly higher values than the other two models), since the star formation rate increases as $Q_g$ decreases -- essentially the gas is compressed under its own weight and forms stars faster.}
	\label{fig:colPerCrit}
\end{figure*}

As more gas arrives at a region of the disk in a marginally unstable state, the surface density is fixed in the profile given by $\Sigma_{GI}$. Since there is a maximum velocity dispersion for a fixed accretion rate, gas is not allowed to accumulate, lest $\sigma \propto\Sigma_{GI}$ exceed this maximum, so the only thing the gas can do is move elsewhere. The gas will then be transported away from where it arrives until it reaches part of the disk which is stable, where it will pile up until that region too becomes unstable. This `wave' of gravitational instability can be seen propagating outwards in Fig. \ref{fig:simpleBal} in both the fiducial model and the model without star formation, until essentially the entire disk is unstable. The equilibrium between GI transport and accretion appears originally at $r \le 7$ kpc both with and without star formation. This location is picked out by the maximum in $\dot{\Sigma}_{acc} / \Sigma_{GI} \propto r \exp(-r/r_{acc})$, i.e. where gas piles up fastest relative to the amount necessary to be gravitationally unstable, which occurs at $r_{acc}$ for a flat rotation curve.

\subsubsection{The fiducial model}
\label{sec:fiducial}
Having examined the simplified models where we disabled GI transport or star formation, we now turn to our fiducial model which includes both. Recalling the surface density distributions shown in Fig. \ref{fig:simpleCol}, it seems that the fiducial model behaves largely like a superposition of the model without star formation and the model without gravitational instability. 

\begin{figure}
	\centering
	\includegraphics[width=8 cm]{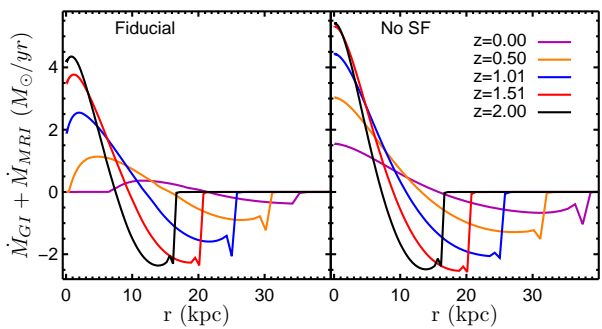} 
	\caption{The inward mass flux through a given radius. Negative values are outward flux. We compare the fiducial model (left) to the same model with star formation turned off (right). As usual, black, red, blue, orange and purple are $z=2.0,1.5,1.0,0.5,0.0$. The outward mass flow is modestly affected by star formation at late times, whereas the inward flow is completely consumed by low redshift. The disk stabilizes in the center, i.e. $\dot{M}\approx 0$, simply because all of the available mass has been consumed by stars. }
	\label{fig:mdot}
\end{figure}

In the previous section we point out that an equilibrium between GI transport and infalling accretion arises when $\Sigma \approx \Sigma_{GI} > \Sigma_{crit}$ (see Fig. \ref{fig:colPerCrit}) and more gas is added. The new gas will be whisked away until it piles up somewhere in the disk that is not yet unstable. If we also include star formation, then rather than being pushed out into a stable region, the gas can be consumed by star formation. Comparing the model without star formation to the fiducial model at $z=2$ and $z=1$ in Fig. \ref{fig:simpleBal}, we can see this effect in action. Gas arrives around $r_{acc}$, and on its way inwards it is consumed by star formation. The balance is then between cosmological accretion and both star formation and GI transport, rather than just GI transport alone. In other words, if the disk can get rid of some gas via star formation, it no longer has to transport it away as fast to maintain $\Sigma \approx \Sigma_{GI}$. Eventually all of the infalling gas at a given radius can be consumed by star formation, and GI transport briefly has no net effect. Just interior to this point though, the cosmological accretion rate is low enough and the star formation rate is fast enough that accretion alone can no longer supply the star formation at that radius, and the stars start forming not from material falling directly onto that radius, but from gas arriving from other parts of the disk via GI transport. This is the point in Fig. \ref{fig:simpleBal} where $\dot{\Sigma}_\mathrm{trans}$ goes from negative to positive. Visually it is clear that the star formation (red) is being supplied by inflowing material (dark blue). 

In this situation, where the star formation is depleting the inflowing gas, the surface density is affected but not necessarily drastically. In steady state, the surface density and velocity dispersion (related via $\Sigma\approx\Sigma_{GI}$) are primarily set by the amount of energy that needs to be dissipated by turbulence, which is set by the amount of torque which must be exerted on the gas to maintain the steady state of matter flowing through the disk at rate $\dot{M}$ (see Fig. \ref{fig:mdot}), which is set by the profile and rate of external accretion. If star formation is removing some of this gas supply, less energy needs to be dissipated and both $\Sigma$ and $\sigma$ will decrease. Eventually, if the star formation rate is fast enough, the inflowing gas (plus the much smaller supply of directly accreting gas) will be entirely depleted and GI will be shut down within that radius. The MRI or some other torque may operate within that radius, and there is certainly still gas within that radius. For $\alpha_{MRI} \la 0.1$, the supply of gas from transport is essentially negligible compared to the supply from continued cosmological accretion. Once the gas supply is shut off in this manner, the gas will burn through the previously $\sim 1/r$ surface density until it reaches equilibrium with the infalling material. At this point newly accreted material is immediately consumed by star formation, and it would take a large burst of accretion to re-activate the GI. In the fiducial model, this shutoff occurs between $z=1$ and $z=0$. For quantitative estimates of when this is important, see section \ref{sec:GIquench}.

The fiducial model also shows a peculiar peak in the star formation rate around $r=17$ kpc at $z=0$ (visible in Fig. \ref{fig:simpleBal} and \ref{fig:tdep}). This corresponds to a peak in the surface density where gas has built up in a ring, which in turn is caused by the fact that the stagnation point in the GI transport flow (i.e. where $\dot{M}=0$) passes through this region. At first gas arrives at this radius from a smaller radius, but at late times it arrives from a larger radius. The location of this stagnation point is set by the boundaries of the GI region, which move outward with time (as a result of GI quenching and the steady viscous spread of the disk), and the particular choice of accretion profile. We therefore expect this feature to exist in many galaxies, but its location and prominence is quite parameter-dependent in our model.

\subsubsection{Energy equilibrium}
\label{sec:energyEq}
\begin{figure*}
	\centering
	\includegraphics[height=17 cm,angle=90]{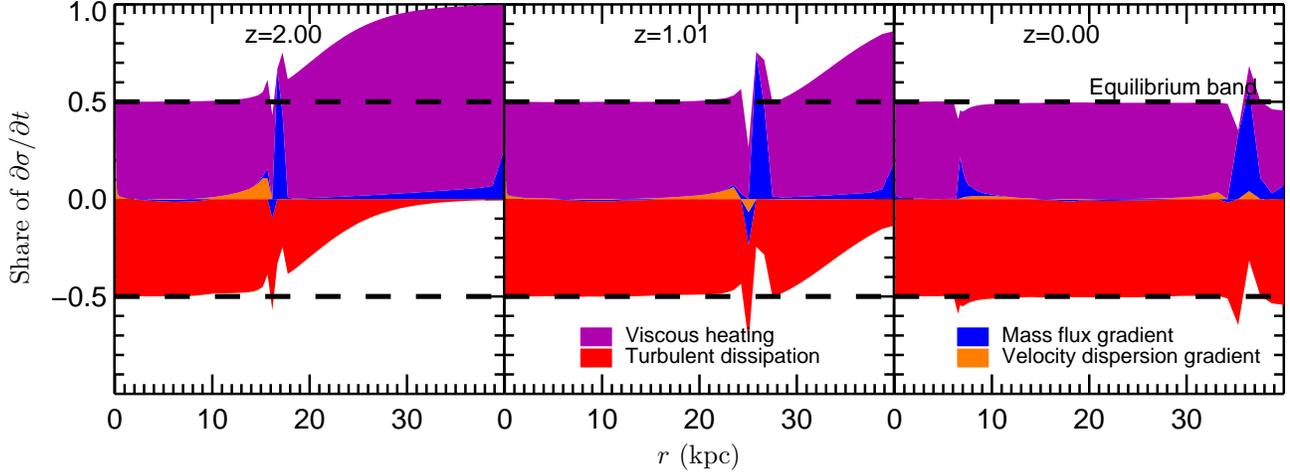} 
	\caption{The balance of terms in the energy equation for the fiducial model. For each term in $\partial \sigma/\partial t$ (equation \ref{eq:dsigdt}), the contribution it makes relative to all the other terms is shown as a function of radius and time. The red region shows cooling $\propto \mathcal{L}$, purple is heating $\propto \mathcal{T}$, orange and blue are terms associated with advection, $\propto \partial\sigma/\partial r$ and $\partial\dot{M}/\partial r$ respectively. Within the regions which are gravitationally unstable, heating and cooling balance almost perfectly. The advection terms are only relevant right where the gravitationally unstable region borders a stable region, where the velocity dispersion and especially the mass flux change dramatically, leading to the spikes near $r=17, 27,$ and $37$ kpc at $z=2, 1$, and 0. }
	\label{fig:balSig}
\end{figure*}
Thus far we have been concerned mostly with the surface density distribution. It is clear that GI transport plays a significant role in setting this surface density. For regions of the disk which are gravitationally unstable, we have asserted that $\Sigma\approx\Sigma_{GI} \propto \sigma/r$. In section \ref{sec:maxSigma} we will show that there is a maximum velocity dispersion set by the mass accretion rate; there is also a minimum velocity dispersion, $\sigma_{th}$ set by the temperature of the gas. This is an adequate first-order understanding of what sets the surface density in the gravitationally-unstable regions, but we have yet to explore what sets $\sigma$ and hence $\Sigma$ between the minimum and maximum values.

Just as with the surface densities, we can show which terms dominate the evolution of $\sigma$ as a function of radius and time (Fig. \ref{fig:balSig}) for the fiducial model. The equilibrium here is even more striking than for the surface densities. Nearly everywhere in the disk, the advection terms (blue and orange) are negligible, and the disk equilibrates between local heating via GI and MRI torques, $\propto \mathcal{T}$ and cooling $\propto \mathcal{L}$ from turbulent dissipation. The exception is at the wave of gas moving outwards to maintain $\Sigma\approx\Sigma_{GI}$ in the inner disk. Here advection becomes important because gas is being transferred from an unstable cell to a stable one with much lower surface density. This stable cell does not pass any mass to the next cell since both have $\mathcal{T} \approx 0$, so $\partial \dot{M}/\partial r$ can be quite large. In reality, the radius separating the gravitationally unstable region from a stable region would be much less well-defined, both because real galaxies are not axisymmetric, and because there may be some `overshoot'. Our model overlooks these effects, so our transition is quite sharp -- a single cell in our simulation. This is the cause of the spikiness, not only in Fig. \ref{fig:balSig}, but also Fig. \ref{fig:simpleBal}, \ref{fig:mdot}, and \ref{fig:distEq}. 

Another exception to the otherwise-good approximation that local heating balances local cooling is at $z=2$ at large radii, where the disk is not gravitationally unstable and the only torque comes from the MRI. This region takes a long time to equilibrate because the dynamical time is quite long, and the MRI is weak, so building up enough turbulent velocity dispersion to be countered by turbulent dissipation takes a few Gyr. Note that this is not the case in the central region at $z=0$ where the disk is again gravitationally stable, but this time the dynamical time is short. Note also that our model implicitly assumes that gas near $\sigma\approx\sigma_{th}$ is in equilibrium between radiative cooling and heating, so the terms we don't show here, e.g. cooling due to metal lines or heating due to the grain photoelectric effect, may dominate in the regions stable to GI. 

Based on Fig. \ref{fig:balSig}, it is safe to approximate the energy balance as entirely local, i.e. to neglect the advection terms, in regions of the disk where GI transport is important. Though our simulations keep all of the relevant terms, we will make this approximation in section \ref{sec:maxSigma} to understand exactly what sets $\sigma$ and $\Sigma$ in gravitationally unstable regions. 

\subsection{Stochastic accretion}

From the previous section, we have seen that a lot depends on the rate of new material being added to the galaxy. This is the term in the continuity equation which increases $\Sigma$, and the disk tends to adjust its available sinks -- star formation (plus galactic winds) and GI transport -- to cancel this out. One may also be concerned that if galaxies do not accrete smoothly at the average rate, the intuition we have built up about a slowly-evolving equilibrium in the previous section may not be applicable to real galaxies. In this section we explore the effect of varying the accretion history stochastically.

\begin{figure*}
	\centering
	\includegraphics[width=17 cm]{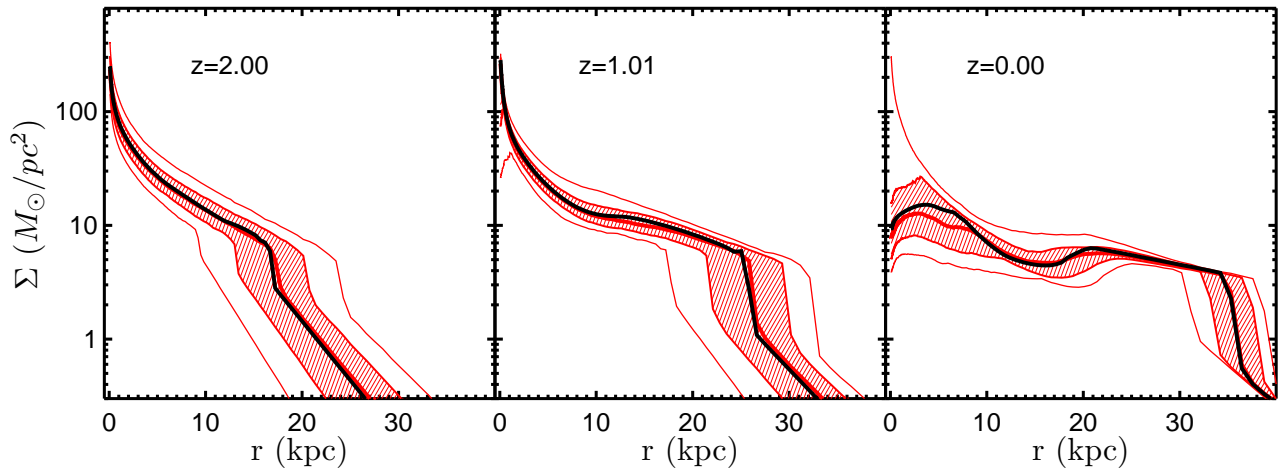} 
	\caption{The radial surface density profile of gas for different redshifts. The black line shows the fiducial model, and the red lines show the median and central 68 per cent (shaded) and 95 per cent of the models with stochastic accretion histories. The variation between surface density profiles at a given radius and time depends mostly on whether the galaxy is gravitationally unstable there. The variation in external accretion rate is largely responsible for the differences between galaxies in regions of the disk which are gravitationally unstable. Additional dependences on parameters of the model and physical properties of the galaxy are shown in appendix \ref{app:vary}. }
	\label{fig:distCol}
\end{figure*}

Fig. \ref{fig:distCol} shows the distribution of surface densities for the same 400 galaxies whose accretion histories were shown in  Fig. \ref{fig:accHistory}, plus the fiducial smooth model for reference. These galaxies all have the same radial scale, namely $r_{acc}=6.9$ kpc. At high redshift the galaxies have similar profiles - $1/r$ profiles at small radii and exponential profiles at large radii. The variation is mostly due to the different gas masses of each galaxy, largely the result of the variation in initial halo mass. Regions of galaxies that are gravitationally unstable have similar $\Sigma \approx \Sigma_{GI}$, since $\Sigma_{GI}$ varies only weakly with accretion rate (see section \ref{sec:maxSigma}). As a consequence, the radii over which the galaxy is gravitationally unstable is just a matter of how far the gas needs to be pushed away from where it arrives to maintain $\Sigma\approx\Sigma_{GI}$.

By low redshift, the galaxies have become remarkably similar at large radii but with more than an order of magnitude variation near the center. At large radii, the disk tends to be gravitationally unstable, but in contrast to the high redshift case, these galaxies all have the same halo mass and so are quite similar in terms of the available gas budget. Meanwhile at small radii, some galaxies, namely those with a recent burst of accretion, are still gravitationally unstable and so exhibit the same $1/r$ profiles seen at high redshift, while others have stabilized and are in an equilibrium between infalling gas and star formation. Thus GI transport greatly magnifies the different accretion rates, causing a wide range of column densities near the center of the galaxy, but at the same time gravitational instability enforces remarkable similarity at large radii.

\begin{figure*}
	\centering
	\includegraphics[width=17 cm]{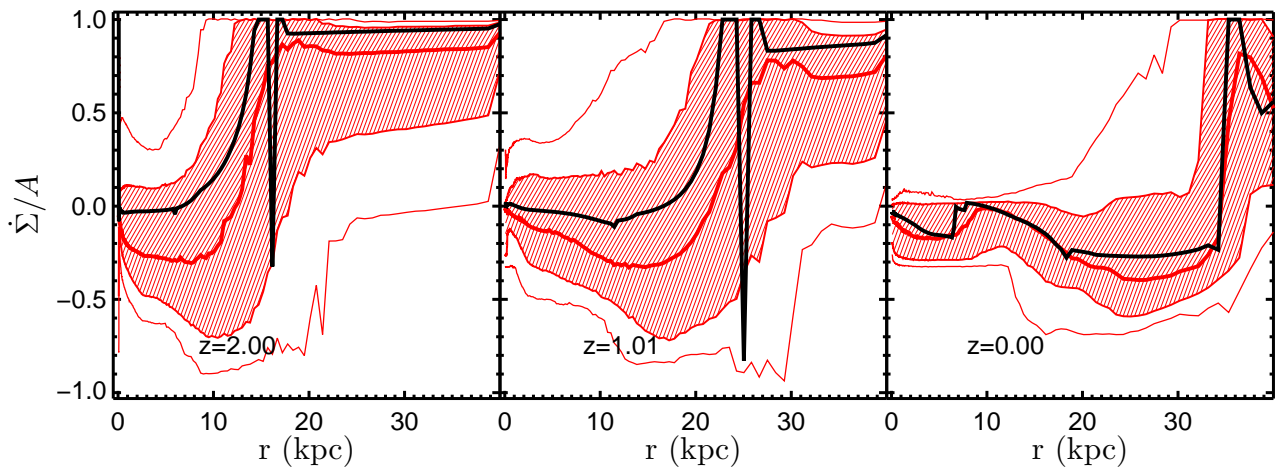} 
	\caption{Inside-out equilibration. Here we show, for the smooth accretion model (black) and the median, central 68 per cent (shaded) and central 95 per cent of the stochastically accreting ensemble of galaxies (red), the radial distribution of $\dot{\Sigma}$ divided by $A$, where $A$ is the sum of the absolute value of each term contributing to $\dot{\Sigma}$. Values of $\dot{\Sigma}/A$ near 1 or -1 indicate that the surface density is changing entirely due to a single term in the equation, while values near zero mean terms of opposing sign are canceling and the surface density is close to equilibrium. Equilibration occurs from inside out, though significant deviations from equilibrium are possible - in fact the typical galaxy is in a low-accretion-rate state and burning through the gas from a past accretion event. Galaxies are also out of equilibrium at large radii where the gas is mostly atomic and hence star formation is slow. }
	\label{fig:distEq}
\end{figure*}

Whether the galaxies are in equilibrium is shown explicitly in Fig. \ref{fig:distEq}.  As with the fiducial model, the ensemble of disks tends to equilibrate from the inside out. The most remarkable difference is the significant fraction of galaxies which are out of equilibrium, not because they are building up gas, but because they are burning through excess gas. These are galaxies which had a burst of accretion followed by a lull. Most galaxies in our stochastic sample are in this state because of the lognormal distribution of accretion rates, which vary on time-scales that are typically short compared to the depletion time. At any given time, a galaxy is therefore likely to be accreting gas slowly but still working through gas that was accreted in a recent burst.

\subsection{Comparison with observations}

Using high resolution and high sensitivity data to infer the HI and ${\rm H}_2$ distributions in nearby spiral galaxies, \citet{Bigiel2012} found that these galaxies have neutral gas surface density profiles well-approximated by a simple exponential,
\begin{equation}
\Sigma_{UP} = 2.1 \Sigma_{tr} e^{-1.65 r / r_{25}}.
\end{equation}
Here $\Sigma_{tr}$ and $r_{25}$ are empirical quantities derived from the data, respectively the surface density at which a particular galaxy has $\Sigma_{HI}=\Sigma_{{\rm H}_2}$ and the radius of the 25 magnitude per square arcsecond B-band isophote. To compare to our simulations, we need to determine these quantities in our own simulated data. We can find $\Sigma_{tr}$ in our simulations by searching for the location where $f_{{\rm H}_2} = 0.5$. In our model this is determined by the \citet{Krumholz2009c} formula, in which this transition surface density is set by the metallicity. The value we should use for $r_{25}$ is somewhat more ambiguous. B-band luminosities are, roughly speaking, set by the star formation rate averaged over at least gigayear time-scales, and the exact luminosity derived for a particular star formation history is somewhat model-dependent. To avoid this issue, we note that if the universal profile is correct, it can be written just as well
\begin{equation}
\Sigma_{UP} = 2.1 \Sigma_{tr} \exp\left( -0.74 r/ r_{tr}  \right)
\end{equation}
where $r_{tr}$ is the radius at which $f_{{\rm H}_2} = 0.5$. This is because $\Sigma_{UP} = \Sigma_{tr}$ at $r=r_{tr}=0.45 r_{25}$. In this way we avoid the modeling uncertainty in converting between a star formation history and a B-band luminosity, and the uncertainty in our star formation prescription at low surface densities, or equivalently the uncertainty in the value of $f_{{\rm H}_2}$.

 \begin{figure}
	\centering
	\includegraphics[width=8 cm]{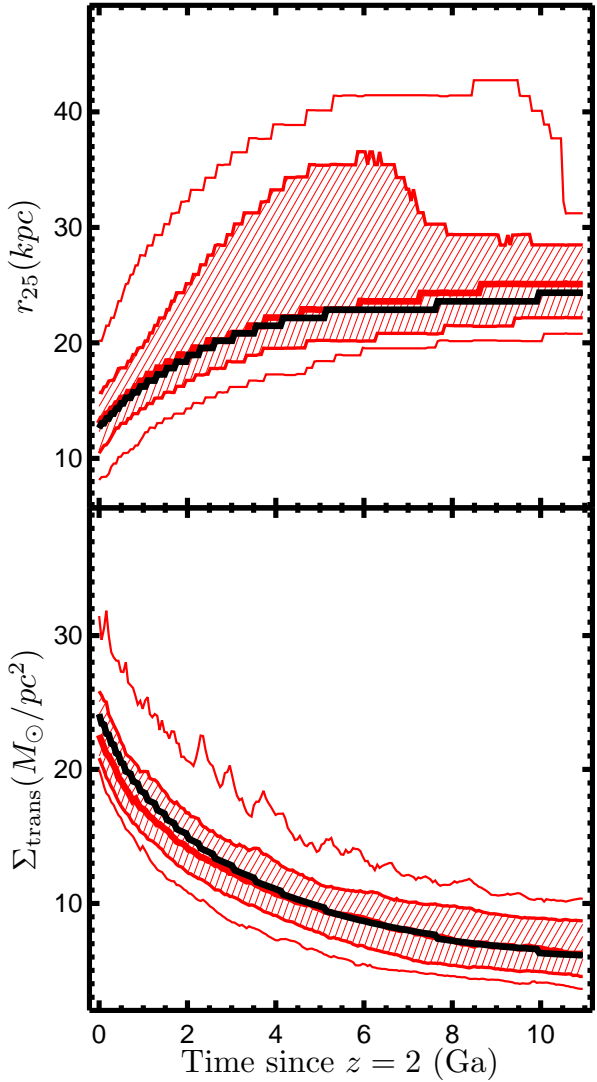} 
	\caption{The parameters that define the universal profile. The median, central 68 per cent (shaded) and 95 per cent of the stochastic ensemble of galaxies are shown in red, with the smooth accretion model (black) for comparison. As the metallicity of the galaxies increases, the column density $\Sigma_{tr}$ at which $f_{{\rm H}_2}=0.5$ falls. As metals build up in the outer disk from local star formation, and advection and diffusion from star formation nearer to the center, the radius at which the molecular-atomic transition occurs, $r_{tr}$, and hence, $r_{25} = r_{tr}/0.45$ steadily increase. }
	\label{fig:distBBparams}
\end{figure}

\begin{figure*}
	\centering
	\includegraphics[width=17 cm]{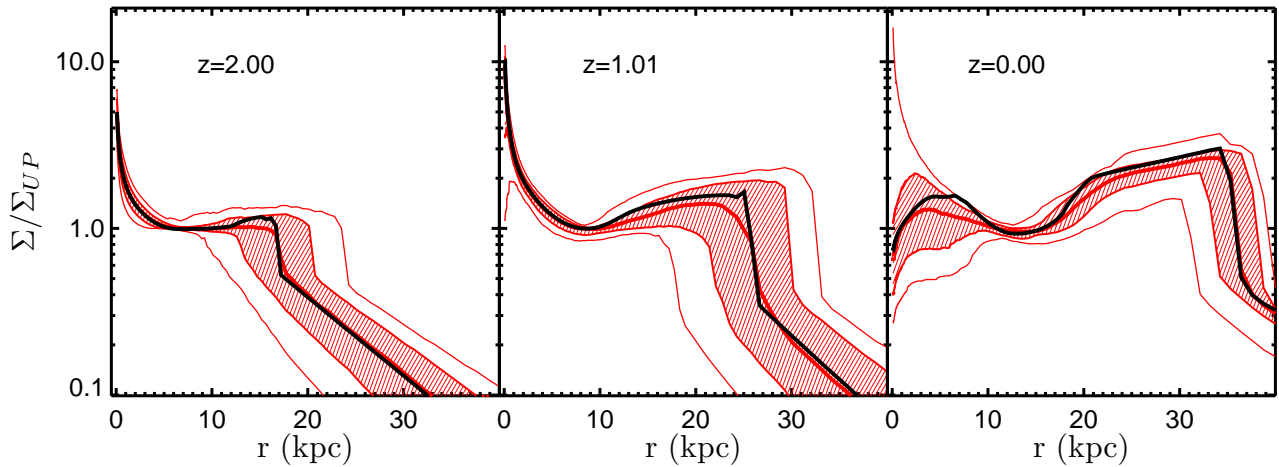} 
	\caption{The ratio of the surface density to the universal profile inferred by measuring $r_{tr}$ and $\Sigma_{tr}$ for each simulation for each time. Black shows the smooth accretion model, while red shows the median, 68 per cent (shaded) and 95 per cent of the distribution for a sample of 400 stochastically accreting galaxies.}
	\label{fig:distBB}
\end{figure*}

For each of our galaxies, we can easily compute $r_{tr}$ and $\Sigma_{tr}$ (Fig. \ref{fig:distBBparams}), each as a function of time, to construct the corresponding $\Sigma_{UP}$ (Fig. \ref{fig:distBB}). The agreement is reasonable, within a factor of two of the empirical relation at $z=0$ for most of the simulated galaxies. At large radii, the effects of photoionization may be important- namely the observations are sensitive only to neutral gas, whereas for the low surface densities $\sim 1 M_\odot\ pc^{-2}$, UV radiation may ionize a significant portion of the gas. As in the observed galaxies, the largest scatter occurs within the central region. We argue that this is a consequence of variations in the accretion histories which allow some galaxies to continue to transport gas to their centers via GI torques, while others have stabilized.

\section{Discussion}
\label{sec:discussion}
One of the striking results of our models is the equilibrium that develops between different terms in the continuity equation. In retrospect this is not surprising, especially near the center of the galaxy, where the star formation time is short and the accretion rate is high. The former allows star formation to quickly adjust to whatever supply of gas is available to it, while high accretion rates mean enough gas can build up to make the disk gravitationally unstable which allows the disk to redistribute the gas and prevent it from piling up wherever it happens to land.

We discuss, roughly in chronological order, or more to the point, in order of decreasing external accretion rate the implications of this slowly evolving equilibrium. At high redshift, the galaxy experiences the maximum surface density it can obtain via an equilibrium between cosmological accretion and GI transport. (section \ref{sec:maxSigma}). GI transport is eventually shut off via star formation (section \ref{sec:GIquench}), after which each annulus near the center of the disk reaches an equilibrium between local gas supply and local star formation (section \ref{sec:SFeq})

\subsection{Maximum velocity dispersion}
\label{sec:maxSigma}

Conservation of angular momentum requires that $\partial \mathcal{T} / \partial r = -\dot{M} v_\phi (1+\beta)$ (equation \eqref{eq:angMom}). At a particular time, we see that the torque at a given radius can be calculated by integrating
\begin{equation}
\mathcal{T}(r) = \mathcal{T}(r=r_0) - \int_{r_0}^r \dot{M}(r') v_\phi(r')(1+\beta(r')) dr'.
\end{equation}
In our numerical model the rotation curve, and hence $v_\phi$ and $\beta$ are fixed in time, as is the inner boundary condition, $\mathcal{T}(r=r_0) = 0$. Thus the torque as a function of radius is exactly mapped to $\dot{M}(r)$. In a steady state, we also know that $\dot{M} < \dot{M}_{ext}$, since otherwise the surface density would be decreasing somewhere to increase it somewhere else. For the moment, we can specialize to a flat rotation curve for which
\begin{equation}
|\mathcal{T}| < \mathcal{T}_{max} \equiv \dot{M}_{ext} v_{circ} r.
\end{equation}
This relation will still hold approximately for somewhat flat rotation curves, since, given the finite supply of new gas $\dot{M}_{ext}$, typically $\dot{M}$ will be significantly less than $\dot{M}_{ext}$ owing to the effects of star formation and outward mass flow, necessary to conserve angular momentum.

\begin{figure}
	\centering
	\includegraphics[width=8 cm]{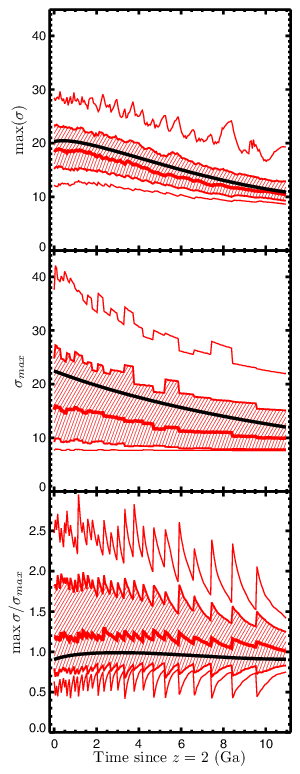} 
	\caption{Simulated vs predicted velocity dispersion. Here we show the maximum value of $\sigma$ measured in our simulations (top), the value of $\sigma_{max}$ predicted by equation \ref{eq:sigMax} (middle panel) and their ratio, $\max (\sigma) / \sigma_{max}$ (bottom). At every change in the accretion rates, the predicted $\sigma_{max}$ jumps and it takes some time for each galaxy to adjust to its new accretion rate. As usual the black line shows the fiducial model and the red lines show the median, central 68 per cent and 95 per cent of the distribution for the stochastically accreting models.}
	\label{fig:distSigmaParams}
\end{figure}

We now employ the assumption of local energy balance, i.e. that the value of $\sigma$ is set by local heating and local cooling with negligible contribution from advection. This assumption is well-satisfied in gravitationally unstable regions of our simulations. Under this assumption,
\begin{equation}
\label{eq:locEnergy}
\frac13 \eta \Sigma \sigma^2 \kappa \left( 1- \sigma_{th}^2/\sigma^2\right)^{3/2} = \frac{(\beta-1) v_\phi}{6\pi r^3 } \mathcal{T}
\end{equation}
Rearranging and approximating $\Sigma \approx \Sigma_{GI}$,
\begin{equation}
\label{eq:torqueBal}
\mathcal{T} = 6 r\eta (\beta+1)v_\phi  \sigma_{th}^3 (\sigma^2/\sigma_{th}^2-1)^{3/2}/((\beta-1)G  Q_{GI})
\end{equation}
Again specializing to a flat rotation curve and defining the dimensionless number $\mathcal{N} \equiv  Q_{GI} G \dot{M}_{ext} / 6\eta \sigma_{th}^3 = 1.8 \dot{M}_{ext}/(1 M_\odot\ yr^{-1})$ and imposing the requirement that $-\mathcal{T} \la \mathcal{T}_{max}$, we arrive at the condition
\begin{equation}
\label{eq:sigMax}
\sigma \la \sigma_{max} \equiv \sigma_{th}  ( \mathcal{N}^{2/3} + 1)^{1/2}
\end{equation}
Thus we see that the velocity dispersions of galactic disks are a direct consequence of cosmological accretion and energy equilibrium. We compare this prediction with the maximum measured values of $\sigma$ in our simulations in Fig. \ref{fig:distSigmaParams}. From the decay of the spikes in the bottom panel, we see that the time-scale to reach the steady state assumed in our derivation can be of order a Gyr. We also see that the central value of $\max(\sigma)/\sigma_{max}$ is remarkably close to unity, meaning that $\sigma_{max}$ is more of an estimate of $\max(\sigma)$ than an upper limit. We note that the measured $\max(\sigma)$ can exceed the predicted maximum slightly even for the smooth accretion model because the assumptions we made in deriving the limit are only approximately true - in particular $\Sigma \approx \Sigma_{GI}$ becomes a worse approximation as the stellar and gaseous velocity dispersions diverge from each other. Meanwhile the stochastic histories are likely to have $\max(\sigma) > \sigma_{max}$. This is because $\sigma_{max}$ depends on the instantaneous accretion rate only, but since the accretion rate changes quickly, the galaxy is likely to still be adjusting to a past burst of accretion.

Even the most extreme galaxies in our population only have $\dot{M}_{ext} \sim 100 M_\odot\ yr^{-1}$, implying $\sigma/\sigma_{th} \la 5.7$, while a more typical $z=2$ galaxy might only have $\dot{M}_{ext} \sim 10 M_\odot\ yr^{-1}$, implying $\sigma/\sigma_{th} \la 2.4$. Since of course $\sigma/\sigma_{th} \ge 1$, the surface density in gravitationally unstable regions can typically only vary by a factor of a few at a fixed radius and $v_{circ}$. We note that the velocity dispersions we show here are somewhat smaller than those observed in the SINS galaxies; however, our MW-progenitor models likely have lower masses than the observed galaxies, and we have included no drivers of turbulence besides gravitational instability.

The exact way that $\sigma$ varies between $\sigma_{th}$ and $\sigma_{max}$ (Fig. \ref{fig:distSig}) depends on the particular accretion profile feeding the galaxy (which roughly determines the shape of the $\sigma$(r) profile), the total amount of gas accreted previously (which sets the outer boundary of the GI region), and star formation (which sets the inner boundary). Qualitatively, the velocity dispersion is highest near the center of the galaxy, since most of the accreted mass arrives near the center of the galaxy and flows inwards. At low redshift this is no longer true because the center of the galaxy becomes gravitationally stable, so the velocity dispersion is forced towards its thermal value $\sigma_{th}$. The outer edge of the unstable region moves outwards as well, since GI transport will always move some gas outwards to conserve angular momentum. This gas is barely touched by star formation given the low molecular fraction at large radii, so over cosmological time that gas will continue to build up and the edge of the gravitationally unstable region will march outwards. Stabilization at small radii and destabilization at large radii lead the whole unstable region to move outwards in time. The lower velocity dispersions in the unstable region, the result of the decreasing cosmological accretion rate, leads to lower characteristic clump masses as estimated by the 2D Jeans mass, $M_J = \sigma^4/G^2\Sigma$, shown in Fig. \ref{fig:distMJ}.

\begin{figure*}
	\centering
	\includegraphics[width=17 cm]{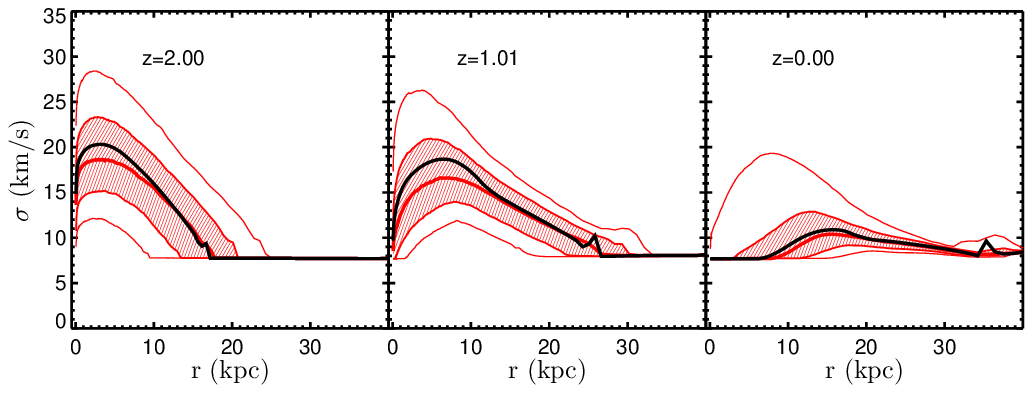} 
	\caption{The velocity dispersion distribution for our models as a function of radius and time. The median, central 68 per cent (shaded) and 95 per cent for the stochastic accretion models are shown in red, with the smooth accretion model in black for reference. The high velocity dispersions at high redshift characterize galaxies undergoing `violent disk instability', which manifests itself most dramatically with giant clumps. At low redshift, the turbulent velocity dispersions driven by GI are much lower and occur further out in the disk, largely as a result of the falling accretion rate. Galaxies undergo this transition, from violent, dynamical evolution to a `secular' evolution, smoothly.}
	\label{fig:distSig}
\end{figure*}

\begin{figure*}
	\centering
	\includegraphics[width=17 cm]{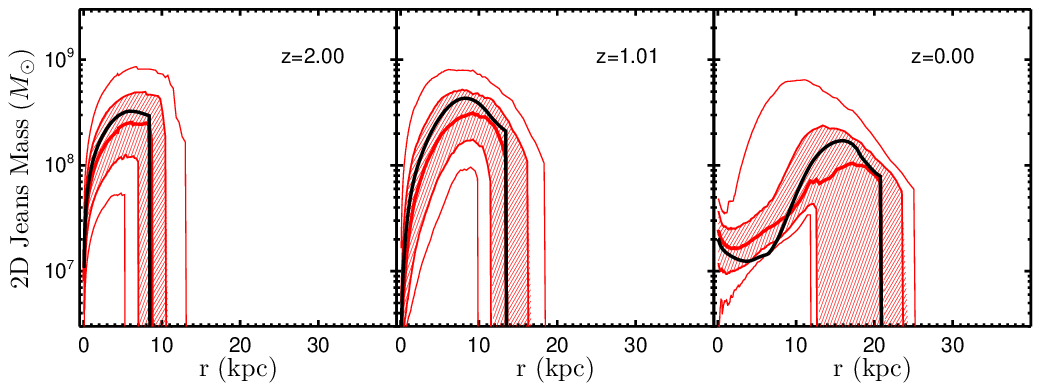} 
	\caption{The characteristic size of clumps in the star-forming disk. Here we show the distribution of the 2D Jeans mass in regions where the molecular fraction is larger than $f_{{\rm H}_2,min}$. The typical mass of gravitationally bound clumps decreases with time, and the peak moves outward in radius. The median, central 68 per cent (shaded) and 95 per cent of the values at each radius and time for the ensemble of stochastically accreting galaxies is shown in red, along with the fiducial model in black.}
	\label{fig:distMJ}
\end{figure*}

The maximum value of $\sigma$ immediately implies a maximum surface density for a flat rotation curve,
\begin{equation}
\Sigma \la \frac{(3/2) v_{circ} \sigma_{th}}{\pi G r Q_{GI}}   ( \mathcal{N}^{2/3} + 1)^{1/2} = \Sigma_{crit} \frac{\sigma_{max}}{\sigma_{th}}.
\end{equation}
Since $\Sigma_{crit}$ for a given model is a fixed function of radius, we immediately see that at a given radius $\Sigma$ in a gravitationally unstable region will also only vary by a factor of a few. However $\Sigma$, unlike $\sigma$, may fall below the value corresponding to $\sigma=\sigma_{th}$. This typically happens because some process has shut off GI transport (section \ref{sec:GIquench}), at which point the disk will equilibrate to a new, lower value of $\Sigma$ (section \ref{sec:SFeq}). We also note that, at least for galactic disks, this maximum column density is likely to be much more restrictive than the one proposed by \citet{Scannapieco2013}, which is based upon the requirement that the rate of turbulent energy dissipation must be removable by radiative cooling.

\subsection{GI quenching}
\label{sec:GIquench}

GI transport shuts off when star formation can consume all of the transported gas. To get an idea of where this happens, we can compare the rate at which a region of the disk, between inner radius $r_A$ and outer radius $r_B$, is resupplied to the rate at which stars are formed within this region.
\begin{equation}
\frac{\dot{M}_\mathrm{supply}}{\dot{M}_{SF}} \approx \frac{\dot{M}(r_B)}{\int_{r_A}^{r_B} 2 \pi r (f_R+\mu)\dot{\Sigma}_*^{SF} dr}
\end{equation}
When this ratio is $\ll 1$, the region in question would easily deplete the gas supply and shut down GI transport, while when it is $\gg 1$, star formation makes no difference and gas flows through the region unharmed. To evaluate this ratio, we use the star formation rate for the Toomre regime, on the grounds that once star formation is slow enough to be in the single-cloud regime, it is unlikely to be hugely important anyway and this ratio will just be $\gg 1$. On similar grounds, we can also assume $f_{{\rm H}_2} \approx 1$, $\Sigma \approx \Sigma_{GI}$  and $Q_g \approx (2/3) Q_{GI}$. In that case, our ratio becomes
\begin{equation}
\frac{\dot{M}_\mathrm{supply}}{\dot{M}_{SF}} \approx \frac{\dot{M}(r_B) G Q_{GI}^2 \pi \left(1 + 2 Q_{GI}/3Q_{lim}\right)^{-1/2}}{\int_{r_A}^{r_B} 36\sqrt{2/3} (f_R+\mu) \epsilon_{\mathrm{ff}} (\beta+1) v_\phi^2 \sigma r^{-1}  dr},
\end{equation}
and we have restricted ourselves to regions where star formation is efficient. In practice this means that $r_B$ can be at most a few kpc. As usual, for simplicity's sake we will specialize to a flat rotation curve, for which we can easily evaluate the integral in the denominator assuming $\sigma \sim \sigma_{max} = const.$, leaving $\int_{r_A}^{r_B} r^{-1} dr = \ln(r_B/r_A)$. Recall that $\sigma_{max}$ depends on the external accretion to roughly the $1/3$ power, so unsurprisingly our ratio will decrease with decreasing $\dot{M}_{ext}$, meaning that all else equal, for a low enough accretion rate the inner region of the disk will be quenched. The logarithmic dependence on $r_B/r_A$ means that in the Toomre regime of star formation, depletion of a fixed gas supply $\dot{M}(r_B)$ is self-similar.

More explicitly, the gas supply is exhausted when $\dot{M}_\mathrm{supply}/\dot{M}_{SF} =1$, which occurs for
\begin{equation}
\label{eq:quenchingRadius}
r_A = r_B \exp\left(- 0.24 v_{220}^{-2}\epsilon_{0.01}^{-1}\frac{1.54}{f_R+\mu}\frac{\dot{M}_{1}}{ \left(1.5 \dot{M}_1^{2/3}+1\right)^{1/2}} \right)
\end{equation}
where we have neglected the additional scalings with $\sqrt{1+Q_g/Q_*}$ and we have introduced a few scaled parameters, $v_{220} = v_{circ}/(220\ \rm{km}\ \rm{s}^{-1})$, $\epsilon_{0.01} = \epsilon_\mathrm{ff}/0.01$, and $\dot{M}_1 = \dot{M}(r_B) / (1 M_\odot\ yr^{-1})$. We caution that this formula is for illustrative purposes only, since $v_\phi$ and $\sigma$ are unlikely to be constant. For these values, it turns out that the exponent is fairly close to zero and so relatively insensitive to the exact values. The exponential evaluates to 0.86, 0.64, and 0.43 for $\dot{M}_1 = 1,\ 4,\ 10$. 

This is actually somewhat surprising, since in our fiducial model Fig. \ref{fig:mdot} shows that the mass flux at a few kpc is near $3 M_\odot\ yr^{-1}$, yet the gas reaches the inner edge of the computational domain at $r=80$ pc easily and GI transport is not shut off until much later. This illustrates the dramatic effect of the rotation curve. The essence of the effect is visible even in equation \ref{eq:quenchingRadius}, namely by the time we reach radii well within the turnover in the rotation curve at $r_b = 3 $ kpc, $v_\phi$ is appreciably smaller than 220 km s$^{-1}$, meaning $r_A/r_B$ should be much smaller. The two powers of $v_\phi$ come from (i) the dynamical time's proportionality to the star formation time - stars form more slowly if the freefall time $\propto r/v_\phi$ is longer, and (ii) the requirement that $Q=Q_{GI}$, which implies $\Sigma \approx \Sigma_{GI} \propto v_\phi$ - lower velocities and hence smaller shear means less gas is required to destabilize the disk. Thus lowering $v_\phi$ decreases both the surface density and the star formation rate for a fixed surface density.

Our simulations use a fixed rotation curve which increases as a powerlaw with index $\beta_0=0.5$ near the center, but galaxies with prominent bulges have what we would term negative values of $\beta_0$, i.e. their rotation curves fall with radius near their centers (see e.g. \citet{Dutton2009b}). As gas approaches the center, it would see higher rotation velocities, which, just the opposite of above, would increase the gas surface density required to maintain GI transport and speed up star formation for fixed gas surface density, hence increasing the GI quenching radius $r_A$. We suggest that this may be a specific physical mechanism for morphological quenching \citep{Martig2009}.  In our estimation, the formation of a bulge acts to quench the innermost regions of the galaxy by shutting off GI transport through the increase in $v_\phi$, but other factors contribute, namely the available supply of gas $\dot{M}$ and the radius at which stars begin to form efficiently in a galaxy, $r_B$. We also note that in our model this quenching is not caused by an increase in $Q$ - the increase in $Q$ and the decrease in SFR are both caused by the shutdown of GI transport.

\subsection{The growth of bulges}

\begin{figure}
	\centering
	\includegraphics[width=8 cm]{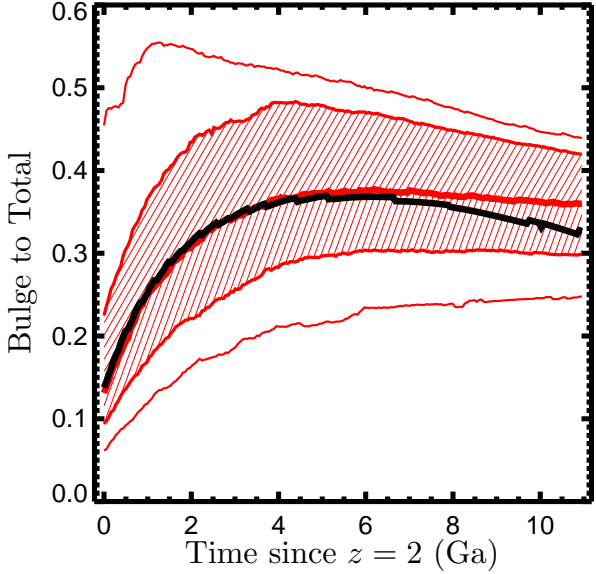} 
	\caption{Estimated bulge to total ratio of the stellar profile. Both the fiducial model (black) and the stochastic ensemble (red) follow similar trends, growing their bulges through GI transport at high redshift, then forming stars preferentially in the disk since $z=1$.}
	\label{fig:distBT}
\end{figure}

Disk instabilities have of late been invoked to explain the growth of spheroids and AGN activity \citep{Dekel2009a,Bournaud2011b,Dekel2013}. Our fiducial choice of parameters certainly funnels gas to the very centers of our model galaxies at a rate of order solar masses per year until $z \sim 0.5$. We caution though that these results depend on our choice of rotation curve, and in particular the rotation curve at the very center of the galaxy. Nonetheless, we can measure the growth of bulges in our simulations. 

There are a number of components which we include in the bulge mass, 
\begin{eqnarray}
\label{eq:bulgeGrowth}
M_B(t) &=& \int_0^t \left(  \dot{M}_*(r=r_0) + \frac{f_R}{\mu+f_R}\left(\dot{M}(r=r_0) \right.\right. \nonumber \\
& & \left.\left. + \int_0^{r_0} 2\pi r \dot{\Sigma}_{cos} dr\right)\right) dt  \nonumber \\
& &+ \int_{r_0}^{r_g} 2\pi r (\Sigma_* - \Sigma_{*,exp}) dr.  
\end{eqnarray}
Starting from $M_B(z=z_{start})=0$, mass enters the bulge a number of ways. First, there is the mass of stars which migrate off the inner boundary of the computational domain $r_0$. Next, there is the gas which does the same, which we assume will quickly form stars. Third, there is gas which, according to our cosmological accretion profile, would accrete within the inner boundary. Lastly, there are stars that are still within the computational domain, but which are in excess of an exponential stellar surface density profile extrapolated inwards from larger radii. We sum all of these components, reducing the gaseous terms by $f_R/(f_R+\mu)$ to account for the fact that for every unit mass of stars formed, only $f_R$ will remain in remnants and $f_R+\mu$ will be lost from the gas supply. The exponential fit $\Sigma_{*,exp}$ is found by 
\begin{equation}
\log \Sigma_{*,exp} = \log \Sigma_*(r_{g}) + r \frac{m\log(\Sigma_*(r_{g}) / \Sigma_*(r_{g}-\Delta r))}{\Delta r} 
\end{equation}
where $r_g = 1.5 r_{acc}(z)$ is the location at which we will fit the local exponential slope, $\Delta r$ is the width of one cell, and $m$ is initially unity. The value of $m$ is gradually reduced until $\Sigma_{*,exp} < \Sigma_*$ at every radius interior to $r_g$ (typically $m=1$ satisfies this condition immediately). This method may overestimate the bulge to total ratio if the stellar profile increases slower than an exponential towards the center, while if the profile is rising faster than an exponential near $r_g$, the contribution to the bulge may be underestimated. The stellar surface density profiles are, however, quite exponential within the star forming region and far from the bulge, likely owing to the mechanism proposed by \citet{Lin1987a}, so this is a reasonable if imperfect estimate. In practice, the flow of gas across the inner boundary, $\dot{M}(r=r_0)$ is the largest of the four terms by a factor of a few, followed by the excess above the exponential.

The growth of bulges measured by the bulge to total (BT) ratio, with the bulge mass estimated by equation \ref{eq:bulgeGrowth}, is shown in Fig. \ref{fig:distBT}. Although gravitational instability funnels gas to the centers of these galaxies, our simulations have star formation efficient enough and a mass loading factor large enough, that the BT ratios tend to lie near $1/3$, a fairly reasonable value for MW-mass galaxies. The trend with redshift seems to be a steep rise between $z=2$ and $z=1$, followed by a very gradual decrease from $z=1$ to $z=0$. This may be attributable to the efficient action of GI at high redshift and its subsequent quenching at lower redshift. Moreover, it is clear that galaxies for which GI transport is important at $z\sim 2$ need not end up as bulge-dominated galaxies at $z=0$. These specific numbers are sensitive to both the angular momentum distribution of infalling gas, and to the parameters which influence star formation, and hence GI quenching, near the center of the galaxy. The galaxies in our sample all have the same accretion scale length at $z=0$, but if we include a 0.4 dex scatter in this parameter, comparable to the scatter in spin parameters observed for dark matter halos in N-body simulations \citep{Bullock2001a}, the central 95 per cent of $z=0$ BT ratios for those galaxies stretches from 0.05 to 0.72.

\subsection{Equilibrium between accretion and SF}
\label{sec:SFeq}
The star formation law has two regimes, so naturally there are two profiles where $\dot{\Sigma}_{cos} = (f_R+\mu) \dot{\Sigma}_*^{SF}$. The simplest case is the single-cloud regime, defined by a constant molecular depletion time $\sim 2$ Gyr. In this regime, 
\begin{equation}
\Sigma = \dot{\Sigma}_{cos} (f_R + \mu)^{-1} \epsilon_\mathrm{ff}^{-1} f_{{\rm H}_2}^{-1} t_{SC}
\end{equation}
This equation is typically not applicable, however, since the outer regions of the disk in the single-cloud regime tend to still be gravitationally unstable even at $z=0$.

Where star formation tends to make a large impact is in the center of the galaxy. In particular, once star formation exhausts the mass flux from GI transport (see the previous section), the supply of gas quickly forms stars until star formation equals the local rate of accretion. This equilibrium is local, in that it occurs independently at each radius, since gas is not being transported between radii. The equilibrium picks out a specific value of $\Sigma$, such that $\dot{\Sigma}_{cos}$ (imposed externally) is roughly equal to $\dot{\Sigma}^{SF}$ (largely determined by $\Sigma$). By the time the disk reaches low redshift, we can assume $Q_g \gg 1$ in this region, so if $f_{{\rm H}_2} \sim 1$ we can calculate that in the central regions of these galaxies,
\begin{eqnarray}
\label{eq:sigmaeq}
\Sigma_{eq}& =&\left(\frac{3\pi\dot{\Sigma}_{cos}^2Q_{lim}\sigma_{th}}{32\epsilon_\mathrm{ff}^2 f_{{\rm H}_2}^2\kappa G (f_R+\mu)^2}\right)^{1/3} \nonumber \\
&\approx &16 \frac{M_\odot}{\rm{pc}^2} \frac{\dot{\Sigma}_{cos,.01}^{2/3}}{(f_R+\mu)^{2/3} f_{{\rm H}_2}^{2/3}}r_1^{1/3}v_{220}^{-1/3}\sigma_{th,8}^{1/3}\epsilon_{.01}^{-2/3}.
\end{eqnarray}
We have used typical values of $\dot{\Sigma}_{cos,.01} = \dot{\Sigma}_{cos}/(0.01\ M_\odot\rm{kpc}^{-2}\rm{yr}^{-1})$, $r_1 = r/(1\ \rm{kpc})$, and $\sigma_{th,8} = \sigma_{th}/(8\ \rm{km}/\rm{s})$. Note that other sources of gas may be added to $\dot{\Sigma}_{cos}$, although if they depend on the star formation rate (e.g., for a galactic fountain) the form of the solution will be a bit different. The assumed accretion rate corresponds closely to the redshift zero value for the smooth accretion history model, and the numerical value of $\Sigma_{eq}$, despite the approximations made, agrees quite well with the simulation. We see that as long as $\dot{\Sigma}_{cos}$ is sufficiently flat, as is the case for an exponential on radial scales much less than the scale length, the value of $\Sigma_{eq}$ will have a moderate increase with radius. This relation will break down if radial transport of gas is operating, and if $f_{{\rm H}_2}$ is appreciably smaller than unity there will be an implicit dependence on $\Sigma_{eq}$ on the right hand side, since $f_{{\rm H}_2}$ is a function of $\Sigma$ (and $Z$).

We saw in section \ref{sec:fiducial} that in our smooth accretion model, the inward mass flux from GI transport is exhausted beginning around $z=0.5$, after which the central gas surface density is rapidly depleted by star formation. We refer to this process as `GI quenching'. When GI transport is active, it essentially collects cosmological infall from all radii and sends most of that gas inwards and some outwards. This can concentrate most of the star formation in the center of the disk, i.e. gas does not form stars at the location it arrives, but in the center of the galaxy. When GI transport is shut off, the center of the galaxy loses this vast supply of gas virtually instantaneously. The surface density falls from $\Sigma \approx \Sigma_{GI} \propto 1/r$ to $\Sigma \approx \Sigma_{eq} \propto r^{1/3}\dot{\Sigma}_{cos}^{2/3}$ in a few depletion times, which may be significantly faster than 1 Gyr (Fig. \ref{fig:tdep}).

We have found that even for large values of an $\alpha$ viscosity (see appendix \ref{app:vary}), and even for a rotation curve quite favorable for transporting gas to the central regions of galaxies, the supply of gas to the central regions of galaxies at $z=0$ via transport through the disk is negligible for a large fraction of the galaxy population. Moreover, gas within this region is unable to move any significant distance radially via these mechanisms. Therefore the equilibrium which develops there is a balance between the {\it local} star formation in some annulus and the {\it local} gas supply. In our model this comes from cosmological infall, but it could in principle also come from supernova-induced accretion \citep{Hobbs2012,Marinacci2010} or gas recycling from old stellar populations \citep{Leitner2011}. Therefore we suggest that measuring the star formation rate and profile in the centers of local galaxies with low star formation rates should directly determine the rate and profile with which those particular regions (regardless of the rest of the galaxy) are being supplied with cold gas.

\section{Summary}
\label{sec:conclusion}

We have explored the evolution of an ensemble of typical disk galaxies with MW-like masses over the past 10 Gyr of cosmic history, with the aim of understanding what sets their surface density profiles.

In our model, disks begin their life at high redshift as exponential and gravitationally unstable in the vicinity of the initial exponential scale length. This is a somewhat artificial initial condition, but by $z=2$ (the simulations are started at $z=2.5$), the  gas has had sufficient time to migrate inwards and the disks become gravitationally unstable interior to the accretion scale length. As more gas is added, the gravitationally unstable region spreads outwards. In this gravitationally unstable state, accreted gas at fairly large radii (of order the accretion scale length) is funneled towards the center of the disk where the high surface densities and short dynamical times allow for efficient star formation. Eventually the cosmological accretion rate falls off and the supply of inflowing gas can be consumed by star formation before the gas reaches the center of the galaxy. At this point the gas transport is shut off and the region of the galaxy interior to this point is quenched, with star formation balancing only the local supply of gas. The main lessons we can draw from these results are as follows:

\begin{enumerate}
\item The surface density at every radius is set by a {\bf slowly evolving equilibrium}. In general, this is a balance among the three terms in the continuity equation: cosmological accretion, star formation, and GI transport. In this paper we have described the properties of the disk when each pair of those terms is in balance. 

\item At a given time, a galaxy will tend to have the following {\bf progression of regions}, from outside inwards. First there is an out-of-equilibrium, low column density region, where gas is building up from cosmological accretion but is not yet gravitationally unstable. Next, the galaxy is in equilibrium between infalling material and GI transport. Further in, star formation takes up an increasing share of the responsibility for balancing incoming accretion - at this point all three terms in the continuity equation are important. Eventually star formation is so efficient that it outstrips the direct supply of gas and can only be balanced by GI transport from larger radii. Finally, if star formation can use up the entire supply of GI-transported gas, there is a quenched region at the center of the galaxy where star formation balances only the direct accretion onto that radius. 

\item If a region is gravitationally unstable, its gas kinetic energy will equilibrate on a dynamical time-scale, with local heating by gravitational instability-driven torques balancing cooling by turbulent dissipation. In a high surface density region where star formation is efficient because of the high molecular fraction and short freefall times, star formation can equilibrate with its gas supply within a few Gyr. The centers of galaxies, where both GI transport operates (at least at high redshift) and stars form efficiently, will therefore generically equilibrate first. Thus {\bf galaxies equilibrate from the inside out}. 

\item In equilibrium, new accretion must be balanced by the available sinks: star formation (plus galactic winds) and transport through the disk. Even at radii where star formation is inefficient, GI transport alone is sufficient to balance accretion. GI transport operates through torques which redistribute angular momentum, allowing gas to be removed from where it accretes. To balance the accretion rate, the gas must lose angular momentum in proportion to the accretion rate, so in steady state the accretion rate specifies the torque. The heating caused by these torques is balanced by turbulent dissipation. The turbulent dissipation rate is proportional to the kinetic energy in the gas, so this balance picks out a velocity dispersion. In summary, the mass flux sets the torque and hence a dissipation rate, which in turn sets the velocity dispersion, so {\bf the cosmological accretion rate sets the velocity dispersion.}

\item In general both the inner and outer boundary of the gravitationally unstable region move outward in time. The inner edge moves outward through a process we call {\bf GI quenching}. As the cosmological infall rate drops, star formation near the center of the galaxy becomes capable of consuming all of the mass moving inwards via GI transport. If all of the gas is consumed on its way in towards the center of the disk, any part of the disk at smaller radii will be deprived of this large supply of gas. Star formation at a particular radius in this quenched region can only be supplied by whatever cold gas is arriving at that particular radius. In our model this is exclusively from direct accretion from the IGM, but there are other plausible sources.  

\item The process of GI quenching becomes more effective at higher rotational velocities, which increase the star formation rate. Massive bulges increase the rotational velocity near the center of a galaxy, so we propose that {\bf morphological quenching} may occur through the following physical channel: GI transport moves gas to the center of a galaxy forming a bulge, the central concentration of matter increases the rotational velocity, GI transport is quenched by the increased star formation rate (and the decreasing cosmological accretion rate), and so the star formation in the central region drops dramatically as its gas supply is removed. The value of $Q$ and $Q_{gas}$ will rise as the gas surface density drops to its new, much lower, equilibrium value. This is distinct from the mechanism proposed by \citet{Martig2009}, wherein they claim that the formation of a spheroid removes the stellar disk and causes the gas disk to stabilize and hence star formation to cease. In our model, the self-gravity of the stars has very little effect on the gas because the stars are assumed to be separately self-regulated to a fixed $Q_*$.  Both models predict a rise in $Q$ and a drop in star formation rate; in our model, both of these are effects of the shutoff in GI transport (which may be hurried by an increased circular velocity from the formation of a bulge), whereas in \citet{Martig2009}, $Q$ increases through the removal of the stars' contribution to the self-gravity to the disk, which then causes the star formation rate to drop.

\item {\bf The growth of bulges} in our simulations occurs primarily through GI transport of gas from the scale on which it is accreted to the centers of galaxies where it forms stars efficiently. Our galaxies all have the same $z=0$ halo mass and accretion scale length, and we recover a relatively narrow range of bulge to total ratios around $0.3-0.4$. If we use a more realistic scatter in accretion scale length of 0.4 dex, the variety of bulge to total ratios increases dramatically.

\item Our simulations show that at $z=0$, some galaxies will be gravitationally unstable at radii $\la 3$ kpc, while others will have undergone GI quenching. The surface density at small radii can therefore vary by an order of magnitude from galaxy to galaxy. This {\bf variability at small radii} is in fact observed in the neutral gas profiles of nearby galaxies studied by \citet{Bigiel2012}. Fundamentally, we predict that this variability is the result of variance in the cosmological accretion rate from galaxy to galaxy, which in turn determines whether the galaxy has undergone GI quenching. Another consequence of this variability is that some galaxies - those which have undergone GI quenching - will have a peak in their star formation surface density in a ring. This may explain so-called `ring galaxies' without invoking a recent merger or bar-induced transport. 

\item The {\bf outer edge of the gravitationally unstable region expands} as more mass falls onto the galaxy. This is because some fraction of the accreted material will move to larger radii until it runs into the edge of the gravitationally unstable region, where it piles up until the disk at that radius also becomes gravitationally unstable. 

\item Although we have emphasized the equilibration of galaxies, we also observe situations where some region of the galaxy is {\bf out of equilibrium}, even in the gravitationally unstable region, and even if the accretion history is perfectly smooth. This occurs primarily in the outer regions of the galaxy where the depletion time and even the dynamical time can be long enough for the cosmological accretion rate to change significantly - in other words the sinks for gas are in equilibrium with a past accretion rate. Equilibrium also breaks down near the moving boundaries between gravitationally stable and unstable regions - for instance when a new region of the galaxy has just lost its gas supply via GI quenching, it takes a depletion time to burn through the (now stationary) gas and reach a new equilibrium with direct accretion. As the surface density of gas, and hence the molecular fraction, decline with time, even the inner parts of the disk may experience depletion times much longer than 2 Gyr, and they too may drop out of equilibrium.

\item The turbulent velocity dispersion of gas in the galaxy falls over time, and the region of the disk subject to GI-driven transport and turbulence moves outwards. This may be interpreted as a {\bf smooth transition from violent to secular instability}. The high velocity dispersions and shorter dynamical times of gas at small radii and high redshift leads to giant clumps (since the Jeans mass $\propto \sigma^4/\Sigma$) evolving rapidly, while at low redshift the gravitationally unstable region has a much longer dynamical time and is characterized by lower clump masses. As predicted in the simpler models of \citet{Cacciato2012}, the violent disk instability which operates at $z=2$ no longer operates today in most MW-mass galaxies, but we show that the transition is gradual and the outskirts of the disk remain unstable even at $z=0$.

\end{enumerate}

We conclude that GI transport is an important driver of disk galaxy evolution. It provides a natural link between MW-like galaxies at the present day and their high-redshift progenitors, and plays a crucial role in determining the structure of disk galaxies.

\section*{Acknowledgments}
The authors would like to thank Eyal Neistein for helpful discussions. JCF is supported by a Graduate Research Fellowship from the NSF. MRK acknowledges support from the Alfred P. Sloan Foundation, from the NSF through CAREER grant AST-0955300, and by NASA through a Chandra Space Telescope Grant and through ATFP grant NNX13AB84G. AB thanks the UCSC astronomy and astrophysics department for its hospitality during his summer visits. AD and AB acknowledge support from GIF grant G-1052-104.7/2009. AD acknowledges support by ISF grant 24/12, and by NSF grant AST-1010033.

\bibliography{mend}

\appendix

\section{Changes since F12}
\label{app:changes}
Here we explicitly list the changes made to our simulation code. In addition to the items discussed below, our code here differs from that of F12 in our assumed rotation curve, assumed accretion rate, the metallicity evolution equation we use, and the star formation prescription we use. These changes are detailed in the main text.
\subsection{Finite volume / explicit mass conservation}
The evolution equations for $\Sigma$ and $\sigma$ are written here in terms of $\mathcal{T}$, $\dot{M}$, and $\partial\dot{M}/\partial r$, as opposed to $\mathcal{T}$, $\partial \mathcal{T}/\partial r$, and $\partial^2 \mathcal{T}/\partial r^2$. The terms involving these quantities are mathematically identical, but this version is clearer physically. Moreover, when we solve these equations, we explicitly calculate the flux $\dot{M}$ from cell ${i+1}$ to $i$, via
\begin{equation}
\dot{M}_{i+1/2} = \frac{-1}{v_\phi(r_{i+1/2})(1+\beta(r_{i+1/2}))}\frac{\mathcal{T}_{i+1}-\mathcal{T}_i}{r_{i+1}-r_i},
\end{equation}
where $i$'s indicate cell-centered quantities and $i+1/2$'s are edge-centered. Using these fluxes, the change in surface density of cell $i$ is then
\begin{equation}
\left(\frac{\partial \Sigma}{\partial t}\right)_\mathrm{transport} = \frac{\dot{M}_{i+1/2} - \dot{M}_{i-1/2}}{2 \pi r_i ( r_{i+1/2} - r_{i-1/2})}
\end{equation}
so that if mass is transported out of cell $i+1$, it must reappear in cell $i$ (or $i+2$). Note that we are using a logarithmic grid, so $r_{i+1/2} = \sqrt{r_i r_{i+1}}$, and $v_\phi$ and $\beta$ may be calculated at these values analytically because of our simple formula for the rotation curve.

The reason this is an improvement is that, written in terms of $\mathcal{T}$ and not $\dot{M}$, $\partial\Sigma/\partial t_\mathrm{transport} \propto \partial^2\mathcal{T}/\partial r^2$. This derivative was computed using a minmod slope limiter, so for example if material attempted to enter or exit a cell from both directions (i.e., the value of $\dot{M}_{i+1/2}$ and $\dot{M}_{i-1/2}$ had opposite signs), $\partial \Sigma/\partial t_\mathrm{transport} = 0$, and so the entering mass would be lost or the exiting mass would remain in the cell. For monotonic solutions of $\mathcal{T}$ this is a small effect, and so only became apparent when mass was added inside the computational domain instead of at its outer edge (see \ref{app:boundary}), which meant some regions would have mass flowing outwards. 

\subsection{Treatment of stable regions where $Q>Q_{GI}$}
In our previous work, when $Q>Q_{GI}$, we solved $f_\mathrm{transport} = 0$. In contrast, in this work we simply set $\mathcal{T}_{GI} = 0$ in those regions. The difference between the two is somewhat subtle. The two treatments would be equivalent if the boundary conditions around the stable region were $\mathcal{T}_{GI,\mathrm{boundary}} = 0$, but this will generally not be the case in our disks because the neighboring unstable regions will have nonzero torques. Thus our previous approach would lead to small but non-zero mass fluxes in stable regions. Our new approach is more consistent with the physical picture we're presenting, namely that radial motion is caused by gravitational instability-induced turbulence.

\subsection{Accretion onto the disk instead of at the outer boundary}
\label{app:boundary}
In our previous work, the accretion of gas onto the galaxy occurred only at the outer boundary of the galaxy, and the accretion rate was enforced by setting $\left(\partial \mathcal{T}_{GI}/\partial r\right)_{r=R} = \dot{M}_{ext}(t) v_{circ}$. We have abandoned this approach because it is inflexible and likely to be physically wrong. In particular, if all the gas comes in at $r=R$, then the value of $R$ may strongly affect the results of the simulation, especially for disks where the accretion rate is not large enough to maintain a gravitational instability, e.g. low mass galaxies or galaxies experiencing a lull in their accretion rate. For these galaxies, accretion at large radii leads to an unphysical pileup of gas in the outermost radial cell. Moreover, the hole in the gas distribution which we saw forming at the center of our simulated galaxy is not a ubiquitous feature in real galaxies, suggesting a more flexible accretion model might be necessary.

In this work, we still need to specify the boundary conditions at inner and outer edges of the computational domain. We opt for the simplest choice, $\mathcal{T}_{GI}(r=r_0) = \mathcal{T}_{GI}(r=R) = 0$, which should be reasonable so long as $R$ is much larger than the radial scale of the accretion.

\section{New Stellar Migration Equations}
\label{app:stmig}
To derive the evolution of stars as a result of their migration through the disk, we will assume that stars obey $d Q_*/dt = \Delta Q_* / T_\mathrm{mig} (2\pi\Omega)^{-1}$, i.e. that stars will exponentially `decay' to a limiting value of $Q_*$ above which they will be stable to gravitational instability, $Q_\mathrm{lim}$, on some multiple of the local orbital time \citep{Sellwood1984,Carlberg1985}. As with the gas, we take the stars to be subject to gravitational torques which will lead to some velocity $v_r^*$ of stars inwards or outwards at each radius such that $Q_*$ approaches $Q_\mathrm{lim}$. In analogy to the gas, we derive evolution equations for the stellar surface density and velocity dispersion which depend on this torque. To do so we begin with the continuity equation and the $\phi-$component of the Jeans equations, both derived from the collisionless Boltzmann equation. The continuity equation is
\begin{equation}
\frac{\partial \rho_*}{\partial t} + \nabla \cdot \left( \rho_* \langle {\bf v}^* \rangle \right) = 0.
\end{equation}
Brackets define an average over the distribution function, namely $\langle v_i^* \rangle \equiv \rho_*^{-1} \int v_i^* f d^3{\bf v^*}$. Note that for simplicity we have taken $f$ to be the distribution function of mass rather than number of stars, where all stars are assumed to have the same mass. The $\phi-$component of the Jeans equations is
\begin{equation}
\frac{\partial \rho_* \langle v_\phi^* \rangle}{\partial t} + \frac{\partial \rho_* \langle v_r^* v_\phi^*\rangle}{\partial r} + \frac{\partial \rho_* \langle v_\phi^* v_z^* \rangle}{\partial z} + \frac{2 \rho_* \langle v_r^* v_\phi^* \rangle}{r} = 0
\end{equation}
As with the gas, we have assumed axisymmetry.

The evolution of the surface density follows almost immediately from integrating the continuity equation in $z$.
\begin{equation}
\frac{\partial}{\partial t} \int_{-\infty}^\infty \rho_* dz = -\frac1r \frac{\partial}{\partial r} \left( r \int_{-\infty}^\infty \rho_* \langle v_r^* \rangle dz\right) - \int_{-\infty}^\infty \frac\partial{\partial z} \rho_* \langle v_z^* \rangle dz
\end{equation}
We will assume that $\langle v_i^* \rangle$ does not vary much over the scale height of the disk, that the disk does not change orientation (so $\langle v_z^* \rangle = 0$), and that $\rho_* \rightarrow 0$ for large and small values of $z$, so integrals over $z$ of the $z$-derivative of a quantity weighted by $\rho_*$ will vanish. Defining $\Sigma_* \equiv \int_{-\infty}^\infty \rho_* dz$, we have
\begin{equation}
\label{eq:starsContinuity}
\frac{\partial \Sigma_*}{\partial t} = -\frac1r\frac\partial{\partial r}\left(r\Sigma_* \langle v_r^* \rangle\right) = \frac1{2\pi r} \frac\partial{\partial r} \dot{M}_*
\end{equation}
where we have defined $\dot{M}_* \equiv -2\pi r \Sigma_* \langle v_r^* \rangle$ to be the inward mass flux of stars through the disk.

To relate this to the torque experienced by the stars, we integrate the $\phi-$component of the Jeans equations in $z$,
\begin{equation}
\frac\partial{\partial t} \Sigma_* \langle v_\phi^* \rangle + \frac1{r^2}\frac\partial{\partial r} r^2 \int \rho_* \langle v_r^* v_\phi^* \rangle dz = 0
\end{equation}
We now define the quantity $\delta v_i \equiv v_i^* - \langle v_i^* \rangle $, the deviation of a particular velocity at a given point from the mean velocity at that point. As usual, we define $\langle \delta v_i \delta v_j \rangle \equiv \sigma_{ij}^2$, and so $\langle v_r^* v_\phi^* \rangle = \langle v_r^* \rangle \langle v_\phi^* \rangle + \sigma_{r\phi}^2$ (since by construction $\langle\delta v_i\rangle=0$). Rearranging, we arrive at
\begin{eqnarray}
\langle v_\phi^* \rangle \frac{\partial \Sigma_*}{\partial t} + \Sigma_* \frac{\partial \langle v_\phi^* \rangle}{\partial t} +   \frac1{r^2}\frac\partial{\partial r} r^2  \Sigma_* \langle v_r^* \rangle \langle v_\phi^* \rangle  = \nonumber \\
 -\frac1{r^2}\frac\partial{\partial r} r^2 \int \rho_* \sigma_{r\phi}^2 dz.
\end{eqnarray}
Using the continuity equation and multiplying through by $2\pi r^2$ yields the evolution equation for specific angular momentum $j_* \equiv r \langle v_\phi^* \rangle$,
\begin{equation}
2\pi r \Sigma_* \frac{\partial j_*}{\partial t} + 2\pi r \Sigma_* \langle v_r^* \rangle \frac{\partial j_*}{\partial r} = \frac\partial{\partial r} \mathcal{T}_*,
\end{equation}
where $\mathcal{T}_* \equiv -2\pi r^2 \int \rho_* \sigma_{r\phi}^2 dz$. As with the gas, we assume a slowly varying potential, in which case we have
\begin{equation}
-\dot{M}_* \frac{\partial j_*}{\partial r} = - \dot{M}_* v_\phi (1+\beta) = \frac{\partial}{\partial r} \mathcal{T}_*,
\end{equation}
so it is clear that the time derivative of $\Sigma_*$ is proportional to the second derivative of the torque. At this point we have also assumed that $\langle v_\phi^*\rangle = v_\phi$, the circular velocity of the gas, so that here, as in e.g. equation \eqref{eq:dsigdt}, $\beta = \partial \ln v_\phi/\partial \ln r$.

To find the evolution of the stellar velocity dispersion, we begin with the collisionless Boltzmann equation,
\begin{equation}
\frac{\partial f}{\partial t} + v_i^* \frac{\partial f}{\partial x_i} - \frac{\partial \psi}{\partial x_i}\frac{\partial f}{\partial v_i^*} = 0
\end{equation}
Next, we multiply through by $v_j^* v_j^*$ and as usual integrate over $d^3 {\bf v}^*$. Since $\psi$, $x_i$, and $t$ are independent of ${\bf v}^*$, we have
\begin{eqnarray}
\int v_j^* v_j^* f d^3 {\bf v}^*  &+& \frac{\partial}{\partial x_i} \int v_i^* v_j^* v_j^* f d^3{\bf v}^* \nonumber \\
&-& \frac{\partial \psi}{\partial x_i}\int v_j^* v_j^* \frac{\partial f}{\partial v_i^*} d^3 {\bf v}^* = 0 
\end{eqnarray}
The final term may be integrated by parts, 
\begin{eqnarray}
\frac{\partial \psi}{\partial x_i}\int v_j^* v_j^* \frac{\partial f}{\partial v_i^*} d^3 {\bf v}^*  &=& -\frac{\partial \psi}{\partial x_i}\int \frac{\partial v_j^* v_j^*}{\partial v_i^*} f d^3 {\bf v}^* \nonumber \\
 &=& -2\frac{\partial \psi}{\partial x_i}\int v_i^* f d^3 {\bf v}^*,
\end{eqnarray}
while the second term may be expanded by again splitting up $v_k^* = \langle v_k^* \rangle + \delta v_k$, so that
\begin{eqnarray}
\langle v_i^* v_j^* v_j^* \rangle &=& \langle v_i^* \rangle \langle v_j^* \rangle \langle v_j^* \rangle +  \langle v_i^* \rangle \sum_j\sigma_{jj}^2 \nonumber \\
& &+ 2 \langle v_j^* \rangle \sigma_{ij}^2 + \rho_*^{-1} \int f \delta v_i \delta v_j \delta v_j d^3 {\bf v}^* 
\end{eqnarray}
For simplicity we drop the final term. This should be a reasonable approximation, since even though the $\delta v_i$ are not necessarily small compared to $\langle v_i^* \rangle$, the integrand contains a quantity which averages to zero, $f \delta v_i$, multiplied by a positive definite quantity $\delta v_j \delta v_j$. We are therefore reweighting an integral which would vanish for a constant weight and approximating it as zero.

With these two substitutions, we arrive at an equation for the evolution of the specific kinetic plus potential energy of the stars,
\begin{eqnarray}
0 &=& \frac\partial{\partial t} \rho_*  \left( \langle {\bf v}^* \rangle^2 + \sum_i \sigma_{ii}^2\right) + \nabla \cdot \rho \langle {\bf v}^* \rangle \left( \langle {\bf v}^* \rangle^2 + \sum_i \sigma_{ii}^2\right) \nonumber \\
 & &+ \nabla \cdot \left(2\rho \langle {\bf v}^* \rangle \cdot \underline{\underline{\sigma^2}}\right) + 2\rho_* \nabla \psi \cdot \langle {\bf v}^* \rangle 
\end{eqnarray}
Here $\underline{\underline{\sigma^2}}$ is the tensor with components $\langle \delta v_i \delta v_j\rangle=\sigma_{ij}^2$. The gravitational work term may be replaced via the continuity equation, since
\begin{eqnarray}
\nabla \cdot (\rho_* \langle {\bf v}^* \rangle \psi) &=& \psi \nabla \cdot (\rho_* \langle {\bf v}^* \rangle) + \rho_* \langle {\bf v}^* \rangle\cdot  \nabla \psi \nonumber \\
 &=& -\frac{\partial (\rho_* \psi )}{\partial t} + \rho_* \frac{\partial \psi}{\partial t}+ \rho_* \langle {\bf v}^* \rangle \cdot \nabla \psi  
\end{eqnarray}
With this substitution, we can group the terms composing the specific energy together, so that if we define $\mathcal{A} = \left( \langle {\bf v}^* \rangle^2 + \sum_i \sigma_{ii}^2 + 2\psi\right)$, we arrive at
\begin{equation}
\frac\partial{\partial t} \rho_* \mathcal{A} + \nabla \cdot \rho \langle {\bf v}^* \rangle \mathcal{A} + \nabla \cdot \left(2\rho \langle {\bf v}^* \rangle \cdot \underline{\underline{\sigma^2}}\right) - 2 \rho_* \frac{\partial \psi}{\partial t} = 0
\end{equation}
Before integrating over $z$, we can use the continuity equation to make one more simplification,
\begin{equation}
\rho_*\frac{\partial}{\partial t}  \left( \mathcal{A} - 2\psi \right)+ \rho_* \langle {\bf v}^* \rangle \cdot \nabla  \mathcal{A} + \nabla \cdot \left(2\rho_* \langle {\bf v}^* \rangle \cdot \underline{\underline{\sigma^2}}\right)  = 0
\end{equation}
Next we approximate $\mathcal{A} \approx v_\phi^2 + \sum_i\sigma_{ii}^2 + 2\psi$, since the other components of $\langle {\bf v}^* \rangle$ are small, $\langle v_r^*\rangle$, or zero, $\langle v_z^*\rangle$. We also approximate $\sigma_{\phi\phi}^2 \approx \sigma_{zz}^2$, in accordance with observations in the solar neighborhood \citep[e.g.][]{Holmberg2009}. Finally, we will again assume that the potential changes slowly, so that $\partial v_\phi/\partial t = 0$. With these approximations, we have
\begin{eqnarray}
0 &=& \rho_* \frac{\partial}{\partial t}\left( \sigma_{rr}^2 + 2\sigma_{zz}^2\right) + \rho_*\langle{v_r^*}\rangle  \frac\partial{\partial r}  \left(v_\phi^2 + \sigma_{rr}^2+2\sigma_{zz}^2 + 2\psi\right) \nonumber \\
& &+ \frac2r \frac\partial{\partial r} r \rho_* \left( \langle v_r^* \rangle \sigma_{rr}^2 + v_\phi \sigma_{r\phi}^2\right) 
\end{eqnarray} 
Now we will integrate over $z$ and assume, consistent with our approximation that $\int f \delta v_i \delta v_j \delta v_j d^3{\bf v} \approx 0$, that $\sigma_{ii}^2$, is roughly constant over a disk scale height. Employing the angular momentum conservation equation and assuming $\partial \sigma_{rr}/\partial t \approx 2 \partial \sigma_{zz}/\partial t$ (J. Sellwood, private communication), we arrive at 
\begin{eqnarray}
\label{eq:dsigrrdt}
\frac{\partial \sigma_{rr}}{\partial t} &=& \frac1{2\pi r \Sigma_*(\sigma_{rr}+\sigma_{zz})} \left(  \frac{v_\phi (\beta-1)}{r^2}\mathcal{T}_* + \sigma_{rr}^2 \frac{\partial \dot{M}_*}{\partial r} \right. \nonumber \\
& & + \left. \dot{M}_* \left(3 \sigma_{rr}\frac{\partial \sigma_{rr}}{\partial r} + 2 \sigma_{zz} \frac{\partial \sigma_{zz}}{\partial r}\right)  \right)
\end{eqnarray}
This is very similar to equation \eqref{eq:dsigdt}, since the procedures used to derive the two are quite similar. The primary distinction is that here we have split the velocity dispersion into a radial and non-radial component whereas for the gas they are assumed to be identical (a reasonable approximation since the gas is collisional). Besides that the only difference is in the numerical values of the coefficients.

With equation \ref{eq:dsigrrdt}, its counterpart for $\partial \sigma_{zz}/\partial t$, and the continuity equation (equation \ref{eq:starsContinuity}) we can follow a similar procedure as detailed in section \ref{sec:gasTransport} to solve for $\mathcal{T}_*$. In particular, we can again split the terms which appear in $dQ_*/dt$ into those which contain $\mathcal{T}_*$ and its radial derivatives, and those which do not.
\begin{eqnarray}
\frac{dQ_*}{dt} &=& \frac{\partial \Sigma_*}{\partial t} \frac{\partial Q_*}{\partial \Sigma_*} + \frac{\partial \sigma_{rr}}{\partial t} \frac{\partial{Q_*}}{\partial \sigma_{rr}} + \frac{\partial \sigma_{zz}}{\partial t} \frac{\partial{Q_*}}{\partial \sigma_{zz}}  \nonumber \\
&=&  f^*_\mathrm{transport}\left(\Sigma_*,\sigma_{rr},\sigma_{zz},\mathcal{T}_*,\frac{\partial \mathcal{T}_*}{\partial r}, \frac{\partial^2 \mathcal{T}_*}{\partial r^2}\right) \nonumber \\
& &+ f^*_\mathrm{source}(\Sigma,\sigma,Z,\Sigma_*,\sigma_{rr},\sigma_{zz})
\end{eqnarray}
None of the equations used so far in this appendix contribute to $f^*_\mathrm{source}$ - the only way $Q_*$ can change without transport is via star formation, which increases $\Sigma_*$ and typically reduces $\sigma_{rr}$ and $\sigma_{zz}$, which in turn tends to lower $Q_*$. For the purposes of computing $\mathcal{T}_*$, we ignore $f^*_\mathrm{source}$ and simply solve $f^*_\mathrm{transport} = \Delta Q_* / T_\mathrm{mig} (2\pi\Omega)^{-1} $ when $Q_*<Q_{lim}$ and set $\mathcal{T}_*=0$ otherwise. This allows $Q_*$ to fall significantly below $Q_{lim}$, though in practice star formation is typically slow enough that $Q_* \approx Q_{lim}$. As with the gas, if this equation yields a solution where $\mathcal{T}_* > 0$, we set $\mathcal{T}_*=0$ in the offending cell.

\section{Sensitivity to parameters}
\label{app:vary}

Thus far we have used only the fiducial parameters, but each one is at least somewhat uncertain (e.g. $\epsilon_\mathrm{ff}$ may vary by a factor of 3 in either direction), or may change for physical reasons (e.g., lower-mass haloes will likely have smaller $r_{acc}$ and $v_{circ}$, and larger $\mu$). To explore the effects of each parameter, we have varied them one at a time from their fiducial values for the smooth accretion history. Fig. \ref{fig:stampColz0} shows the $z=0$ surface density distribution when each parameter is varied. Essentially, all of the models are gravitationally unstable over a wide range of radii and parameter choices. We also show the metallicity distribution for all of these models in Fig. \ref{fig:stampz0Z}. The metallicities are in general hugely sensitive to the parameters, so much so that any attempt to draw a physical conclusion by fitting a metallicity gradient should be treated with extreme caution, since the same metallicity gradient can be produced by varying any number of parameters. We discuss each of the parameters in more detail in the following sections.

\begin{figure*}
	\centering
	\includegraphics[height=17 cm,angle=90]{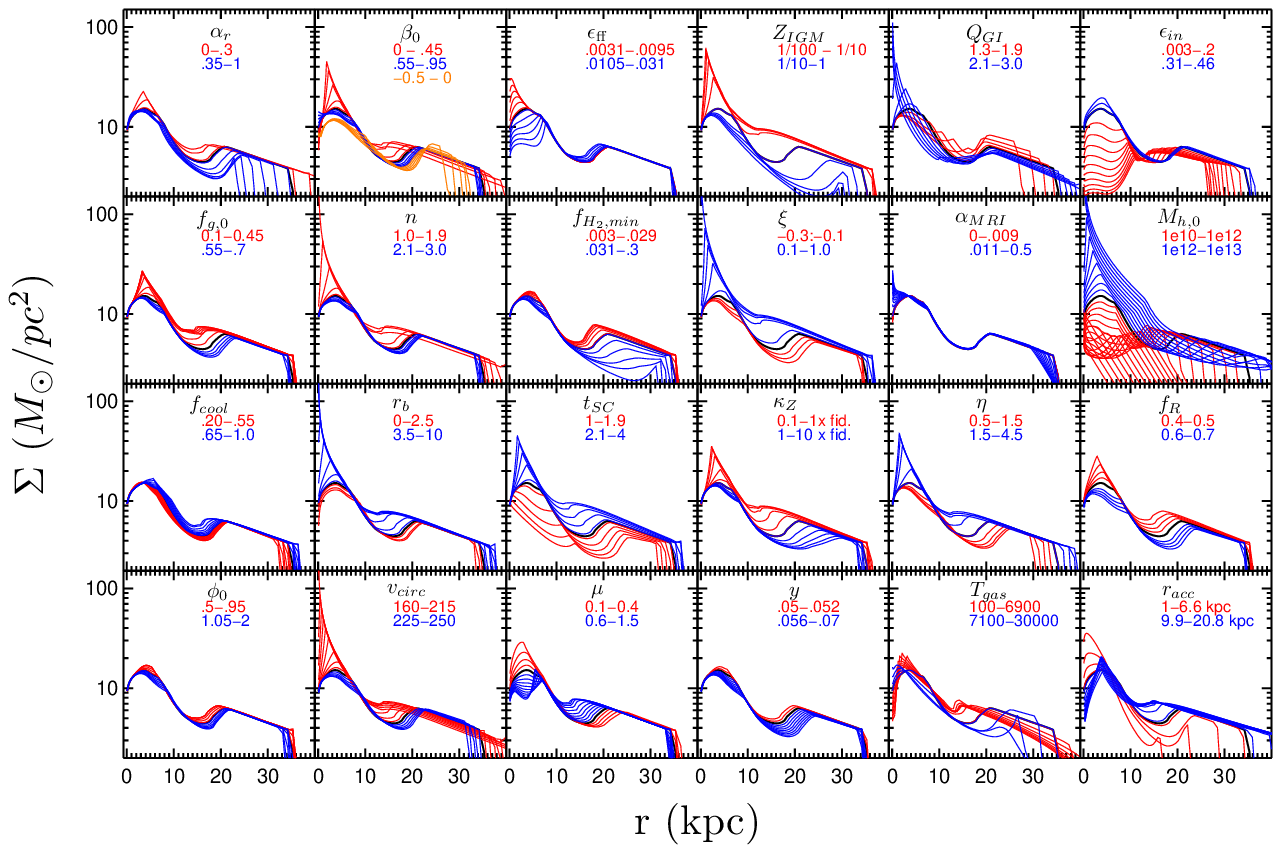} 
	\caption{ Surface density at $z=0$. Each pane shows models where the given parameter is varied within the quoted range - red models have lower values of the parameter, blue higher. The models are arranged in 6 columns according to what the parameters are controlling- from left to right: initial conditions, rotation curve, star formation, metallicity, gas transport, and gas supply.}
	\label{fig:stampColz0}
\end{figure*}

\begin{figure*}
	\centering
	\includegraphics[height=17 cm,angle=90]{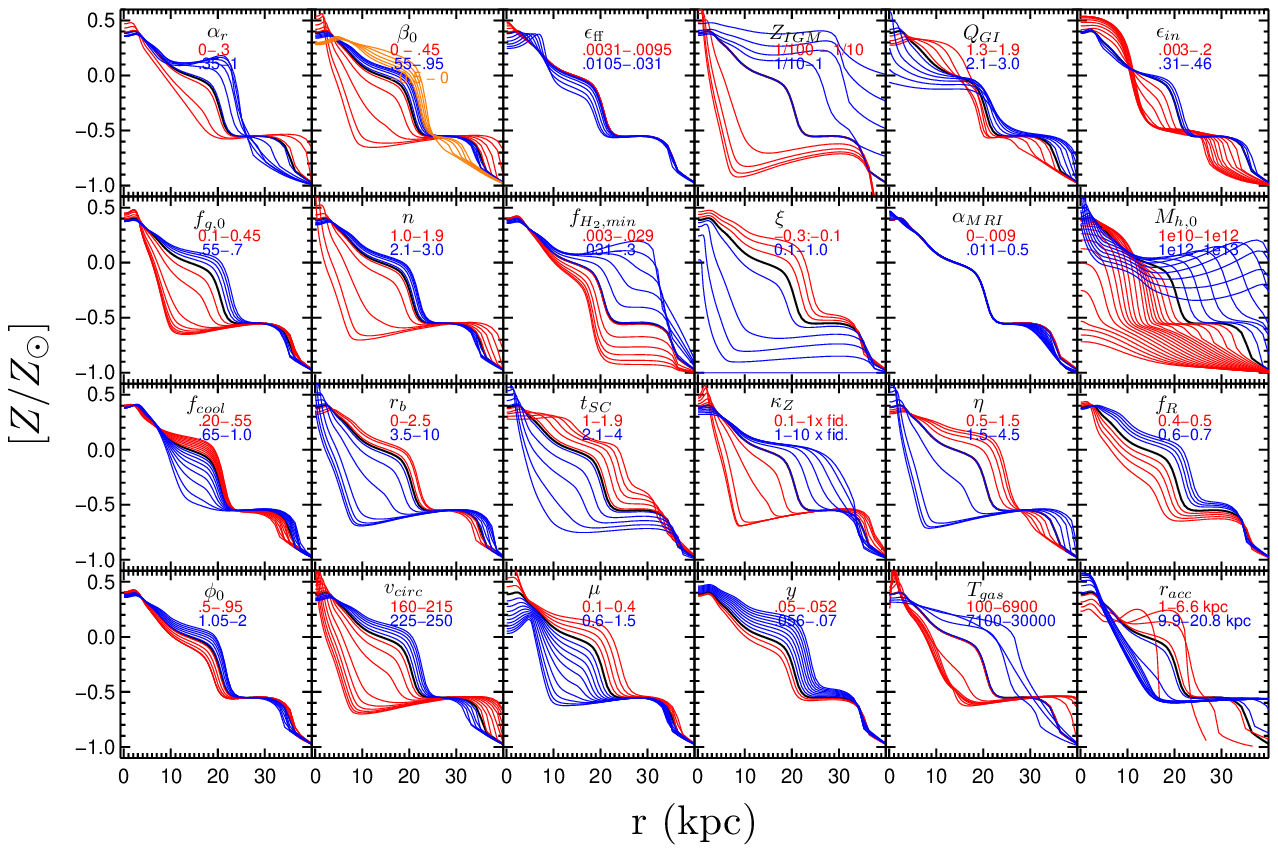} 
	\caption{Metallicity for the same models as in Fig. \ref{fig:stampColz0} }
	\label{fig:stampz0Z}
\end{figure*}

\subsection{Initial conditions  - $\alpha_r$, $f_{g,0}$, $f_{cool}$, $\phi_0$}
These parameters, the scaling of the accretion scale length with halo mass, the initial gas fraction, the fraction of baryons which have cooled into a disk at $z=z_{start}$, and the initial ratio of stellar to gas velocity dispersion, are almost completely irrelevant for the $z=0$ surface density distribution. In our framework, the surface density at each radius is set by an equilibrium relation, and so it is unsurprising that the initial conditions are washed out. The exception is that strong evolution of $r_{acc}$ with halo mass, i.e. (probably unrealistically) high values of $\alpha_r$, lead to smaller gravitationally unstable regions at $z=0$, since so little mass was accreted directly at large radius.

\subsection{Rotation curve - $\beta_0$, $n$, $r_b$, $v_{circ}$}
The shape of the rotation curve is controlled by these four parameters - inner powerlaw slope, the sharpness of, and location of the turnover from flat, and the overall normalization. The most dramatic effect of changing these parameters is in the inner region of the disk, where different values can change the surface density by an order of magnitude. This is because the rotation curve influences both the surface density in gravitationally unstable regions and the star formation rate in the central region, where $\dot{\Sigma}_*^{SF} \propto \kappa \propto v_{\phi}$, and hence has a strong influence on where exactly the disk is able to form stars fast enough to shut off GI transport to the innermost region. Negative values of $\beta_0$ have an effect at large radii too. In general, however, the qualitative behavior of our models is largely insensitive to these parameters except near galactic centers.

\subsection{Star formation - $\epsilon_\mathrm{ff}$, $f_{{\rm H}_2,min}$, $t_{SC}$, $\mu$}
Each of these parameters governs the rate at which gas is depleted from the galaxy, either into stars or galactic winds. As we would expect, $\epsilon_\mathrm{ff}$ is important in the inner region of the disk where the star formation rate is in the Toomre regime, while $f_{{\rm H}_2,min}$ is important in the outer disk where the gas is mostly atomic and hence $SFR \propto f_{{\rm H}_2} = f_{{\rm H}_2,min}$. Note however that the factor of three variation in $\epsilon_\mathrm{ff}$ has a much larger effect than the order of magnitude variation in $f_{{\rm H}_2,min}$. This is because the surface density is set by different equilibria - in the outer disk the surface density is mostly set by gravitational instability, whereas in the inner region the surface density is determined by whether star formation has shut off GI transport to the central region or not, which in turn depends strongly on the star formation law there. If GI transport has been shut off, then the surface density is set by cosmological infall balancing star formation, so a change in the star formation law given a fixed infall rate $\dot{\Sigma}_{cos}$ can have a large effect. The mass loading factor $\mu$ affects the rate of mass loss everywhere in the disk, but again because of the different equilibria, it has a much stronger effect in the inner region. Again, however, we note that the qualitative results, as opposed to the precise numerical values of $\Sigma(r)$, are insensitive to these parameters.

\subsection{Metallicity - $Z_{IGM}$, $\xi$, $\kappa_Z$, $y$}
The metallicity of the infalling and initial gas in the disk, the metal enhancement of galactic winds, the metal diffusion coefficient, and the yield. The first three strongly influence the metallicity of the disk, as does the yield to a lesser extent. This in turn affects the ${\rm H}_2$ fraction when the gas is near its transition surface density (higher (lower) surface densities will have $f_{{\rm H}_2}\sim 1$ ($f_{{\rm H}_2,min}$) regardless. The ${\rm H}_2$ fraction then has an effect on the star formation rate. In general, changes in the parameters which decrease the overall metallicity increase the surface density everywhere by decreasing the rate at which star formation is depleting/ejecting the gas.

\subsection{Transport - $Q_{GI}$, $\alpha_{MRI}$, $\eta$, $T_{gas}$}
The parameters which control the radial transport of the gas have the potential to strongly affect the surface density, since much of the disk is gravitationally unstable. $Q_{GI}$ and $T_{gas}$ both directly affect $\Sigma_{crit}$, namely the minimum surface density for the gas to be gravitationally unstable. Meanwhile $\eta$ only affects the energy balance in the disk. For all the parameters, the primary difference is in where the GI transport is shut off. Higher $Q_{GI}$ and dissipation rate allow the gas to reach farther towards the center before being consumed by star formation. $T_{gas}$ turns out not to matter all that much, primarily because in the limit of large accretion rates / low thermal velocity dispersions, the energy balance is independent of $T_{gas}$. Perhaps the most dramatic parameter here is $\alpha_{MRI}$, which has even less effect on the surface density distribution than the initial conditions.

\subsection{Gas supply - $\epsilon_{in}$, $M_{h,0}$, $f_R$, $r_{acc}$}
Here we come to the most important parameters in setting the surface density - the quantity and distribution of the gas supply. These parameters respectively are the efficiency at $z=0$ (assuming a fixed efficiency at $z=2$), the halo mass at redshift zero (where we change only $M_h$ and hence the accretion history, but no other parameters), the remnant fraction, and finally the accretion scale length. These models obey the trends one might expect. Less gas means the region over which the gas is gravitationally unstable is smaller. Parts of the disk beyond this radius retain the exponential character of the accretion profile, and parts of the disk interior have their surface densities set by the balance between local accretion and local star formation.

\clearpage

    \fi

\end{document}